\documentclass[journal]{./TeX/IEEEtran}

\ifCLASSINFOpdf
  \usepackage[pdftex]{graphicx}
\else
\fi

\usepackage[colorlinks]{hyperref}
\usepackage{multirow}
\usepackage{times}
\usepackage{amsmath,amssymb}
\usepackage{algorithm}
\usepackage{epstopdf}
\usepackage{subfigure}
\usepackage[dvipsnames]{xcolor}
\usepackage{algorithm}
\usepackage{algorithmic}
\usepackage{cite}

\usepackage{bbm}
\newtheorem{theorem}{Theorem}

\newtheorem{definition}{Definition}

\newenvironment{proof}{\paragraph*{Proof}}{\hfill$\square$}

{\begin{list}               
    {$\bullet$ \hfill}{
        \setlength{\leftmargin}{\parindent}
        \setlength{\parsep}{0.04\baselineskip}
        \setlength{\itemsep}{0.5\parsep}
        \setlength{\labelwidth}{\leftmargin}
        \setlength{\labelsep}{0em}}
    }
{\end{list}}

\providecommand{\eref}[1]{\eqref{#1}}  
\providecommand{\cref}[1]{Chapter~\ref{#1}}

\providecommand{\fref}[1]{Figure~\ref{#1}}
\providecommand{\tref}[1]{Table~\ref{#1}}

\providecommand{\R}{\ensuremath{\mathbb{R}}}

\providecommand{\E}{\ensuremath{\mathbb{E}}}

\providecommand{\Pb}{\ensuremath{\mathbb{P}}}

\providecommand{\bydef}{\overset{\text{def}}{=}}

\renewcommand{\vec}[1]{\ensuremath{\boldsymbol{#1}}}
\providecommand{\mat}[1]{\ensuremath{\boldsymbol{#1}}}

\providecommand{\calL}{\mathcal{L}}

\providecommand{\calN}{\mathcal{N}}

\providecommand{\mI}{\mat{I}}

\providecommand{\mK}{\mat{K}}

\providecommand{\mY}{\mat{Y}}
\providecommand{\mZ}{\mat{Z}}

\providecommand{\vw}{\vec{w}}

\providecommand{\veta}{\vec{\eta}}
\providecommand{\vtheta}{\vec{\theta}}

\providecommand{\vlambda}{\vec{\lambda}}

\providecommand{\lambdahat}{\widehat{\lambda}}

\providecommand{\vlambdahat}{\boldsymbol{\widehat{\lambda}}}

\providecommand{\vone}{\vec{1}}

\providecommand{\Var}{\mathrm{Var}}

\newcommand{\round}[1]{\left \lceil #1 \right \rfloor}

\newcommand{\maximize}[1]{\mathop{\underset{#1}{\mathrm{maximize}}}}

\providecommand{\SNRH}{\mbox{SNR}_{\text{H}}}

\begin{document}

\title{HDR Imaging with Quanta Image Sensors: Theoretical Limits and Optimal Reconstruction}

\author{Abhiram~Gnanasambandam,~\IEEEmembership{Student~Member,~IEEE}, and~Stanley~H.~Chan,~\IEEEmembership{Senior~Member,~IEEE}
\thanks{A. Gnanasambandam and S. Chan are with the School of Electrical and Computer
Engineering, Purdue University, West Lafayette, IN 47907, USA. Email: {
\{agnanasa, stanchan\}}@purdue.edu. This work is supported, in part, by the National Science Foundation under grant CCF-1718007.} 
}

\maketitle

\begin{abstract}
High dynamic range (HDR) imaging is one of the biggest achievements in modern photography. Traditional solutions to HDR imaging are designed for and applied to CMOS image sensors (CIS). However, the mainstream one-micron CIS cameras today generally have a high read noise and low frame-rate. Consequently, these sensors have limited acquisition speed, making the cameras slow in the HDR mode. In this paper, we propose a new computational photography technique for HDR imaging. Recognizing the limitations of CIS, we use the Quanta Image Sensors (QIS) to trade spatial-temporal resolution with bit-depth. QIS are single-photon image sensors that have comparable pixel pitch to CIS but substantially lower dark current and read noise. We provide a complete theoretical characterization of the sensor in the context of HDR imaging, by proving the fundamental limits in the dynamic range that QIS can offer and its trade-offs with noise and speed. In addition, we derive an optimal reconstruction algorithm for single-bit and multi-bit QIS. Our algorithm is theoretically optimal for \emph{all} linear reconstruction schemes based on exposure bracketing. Experimental results confirm the validity of the theory and algorithm, based on synthetic and real QIS data.
\end{abstract}

\begin{IEEEkeywords}QIS, high dynamic range, signal-to-noise ratio, photon counting
\end{IEEEkeywords}

\section{Introduction}
Quanta image sensors (QIS) are a new type of image sensors with single photon sensitivity. Originally proposed in 2005, the sensor was designed to overcome the full-well capacity limit of the mainstream CMOS image sensors (CIS) due to the shrinking pixel sizes \cite{fossum200611,fossum2005gigapixel}. Compared to CIS, the anticipated QIS will have a smaller pixel pitch, higher frame rate, smaller read noise, and lower dark current. As recently reported in \cite{Ma:17}, the latest QIS prototype has achieved a read noise below $0.25e^-$ r.m.s. at room temperature and a frame rate beyond 1000 frames per second, while the pixel pitch is only $1.1 \mu$m. The small pixel pitch of the sensors allows higher spatial resolution per unit area. The small read noise and dark current also allow the sensors to perform single photon counting at room temperature. Besides, the fabrication of QIS is based on the standard 3D stacking techniques widely used for CIS \cite{Ma:17}, and hence the potential cost of QIS can be comparable to a CIS. These factors have made QIS an excellent candidate for imaging dark and dynamic scenes. Recent reports in computer vision have proved the effectiveness of the sensors in various applications \cite{gyongy2018single, gnanasambandam2019megapixel,gnanasambandam_chan_eccv2020,chi_chan_eccv2020}.

\begin{figure*}[!]
    \centering
    \includegraphics[width=0.9\linewidth]{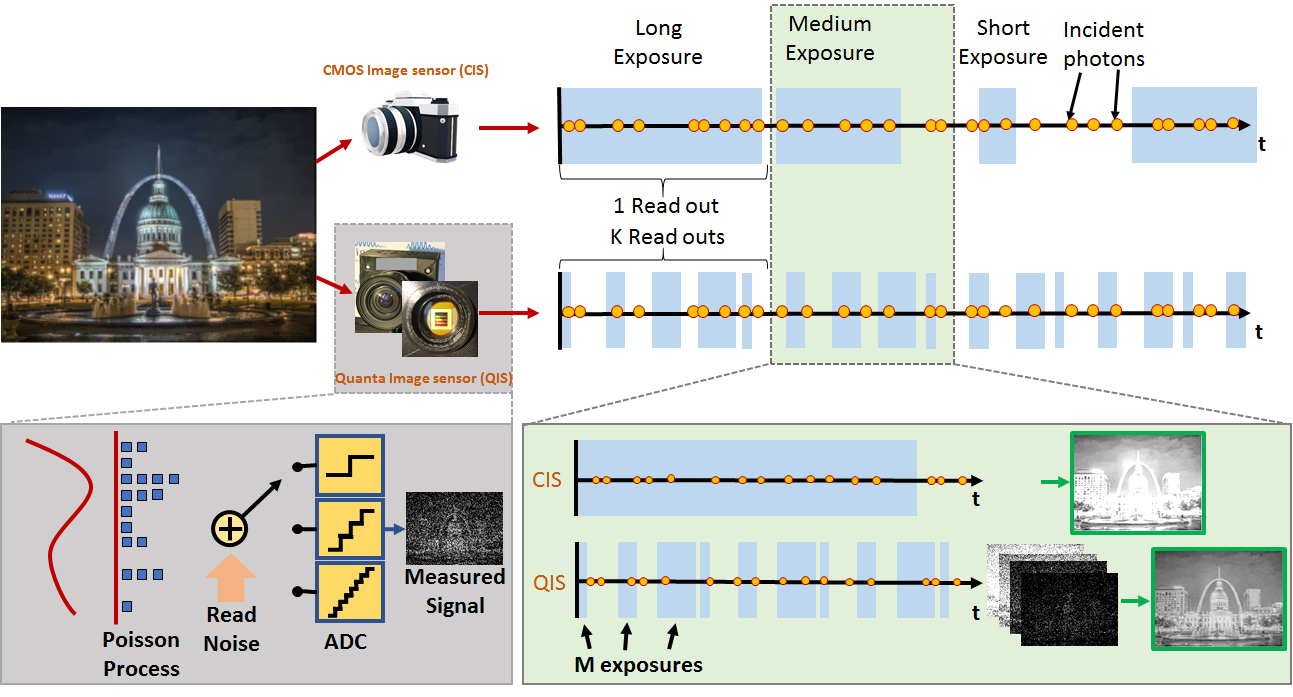}
    \caption{\textbf{HDR Imaging with Quanta Image Sensors.} Quanta Image Sensors have the ability to oversample the scene, because of their significantly higher frame rate. In this paper, we show that when we combine the oversampling ability of Quanta Image Sensors with exposure bracketing, the dynamic range achieved by the system far exceeds the dynamic range of the CMOS Image Sensors. }
    \label{fig:idea}
\end{figure*}

\subsection{Current state-of-the-art HDR imaging}

High dynamic range (HDR) imaging refers to capturing an image where the photon flux varies substantially between different parts of the image. Without a customized acquisition scheme and reconstruction technique, pixels of a HDR scene will saturate in the bright regions whereas pixels in dark regions will not have enough photons. In either case, the signal-to-noise ratio (SNR) will be poor. The goal of HDR imaging is to capture the scene such that we can maintain a consistent signal-to-noise ratio throughout the image.

HDR imaging is ubiquitous in photography, microscopy, navigation, and surveillance, to just name a few. We refer the readers to the texts by Banterle et al. \cite{banterleadvanced} and Reinhard et al. \cite{reinhard2010high} for an introduction to the subject. In general, HDR imaging techniques can be categorized into three families: (i) exposure bracketing \cite{mitsunaga1999radiometric,gallo2016stack,granados2010optimal,debevec2008recovering,sen2016practical}, (ii) coded exposure \cite{serrano2016convolutional,nayar2000high}, and (iii) burst photography \cite{hasinoff2016burst,buades2009note,joshi2010seeing}.

\textbf{CIS-based exposure bracketing}.
In exposure bracketing, we take multiple exposures of the scene, some with longer and some with shorter exposures. Then, we use a carefully designed image processing algorithm to merge these differently exposed images to form the final image. Exposure bracketing is popular because of the simplicity which allows it to be used on hand-held devices. The downside, however, is that it requires capturing many long and short exposure frames before the fusion step. The overall acquisition time is thus long.

Despite the  variety of exposure bracketing techniques, one thing that remains unchanged is the original linear combination idea.  There are multiple ways of achieving linear reconstruction. One can combine the processed images instead of the raw image \cite{debevec2008recovering,mitsunaga1999radiometric,tsin2001statistical}, or directly use the raw data \cite{robertson2003estimation,kirk2006noise,granados2010optimal,kronander2013unified}. The choice of the combination weight also plays a critical role in HDR reconstruction. In \cite{mitsunaga1999radiometric}, the weights are chosen to be proportional to the SNR so that the overall SNR  of the combined image is optimized. Hasinoff et al. \cite{hasinoff2010noise} propose two different ways to obtain an HDR image, either by maximizing the minimum SNR in the image or minimizing the overall time taken to obtain an image with a target SNR. In \cite{robertson2003estimation}, Robertson et al. proved that the maximum likelihood estimate of the HDR estimate is the linear combination with weights inversely proportional to the variance of the signal. Granados et al. \cite{granados2010optimal} extend the work of \cite{robertson2003estimation} by including different sources of noise in the model. Mertens at al. \cite{mertens2009exposure} combine LDR images without converting them into HDR values. Recently, several neural-network based HDR reconstruction methods that hallucinate HDR images from LDR images \cite{eilertsen2017hdr,marnerides2018expandnet} or use exposure bracketed images for reconstructing HDR dynamic scenes have also been proposed \cite{kalantari2017deep, wu2018deep}.

\textbf{CIS-based coded exposure}. Coded exposure can be thought of as a modified version of exposure bracketing, where instead of capturing multiple frames with different exposures, we capture a single frame with different exposures, and use carefully developed algorithms to combine these different exposures into a single image. Some representative works include \cite{nayar2000high}, which proposed using spatially varying exposures to obtain HDR imaging in a single frame, and \cite{serrano2016convolutional} which extended the idea by using convolutional sparse coding.

\textbf{CIS-based burst photography}. The idea of burst photography is to acquire a burst of short exposure frames so that all the frames are below the saturation limit. Then, by properly aligning the images (with respect to object motion and camera motion), one can reconstruct an HDR image. Over the past few years, various burst photography algorithms are proposed, ranging from traditional motion alignment methods \cite{cai2009blind,zhang2014multi,delbracio2015burst} to end-to-end deep learning methods \cite{wieschollek2017learning,aittala2018burst,tao2018scale}. However, CIS-based burst photography is intrinsically limited by the photon sensitivity of the sensors. As the exposure becomes short, the high read noise and dark current of the sensor will prohibit the precise measurement of the signals. This, when added to the random Poisson statistics of the photon arrivals, poses a fundamental limit to CIS-based burst photography which is difficult to be solved by image processing including deep learning algorithms.

\subsection{Quanta Image Sensors}
Being a single-photon image sensor, QIS have substantially better photon sensitivity than CIS. This unique capability opens the door to a new way of acquiring HDR images. In particular, with QIS one can operate in a regime that allows significantly shorter exposure and higher frame rate while still being able to resolve the incoming photons. CIS-based techniques are harder to operate in this regime because as the exposures become short, the read noise and dark current of the sensor will bury detectable signals.

\textbf{Image reconstruction for Quanta Image Sensors.}
In recent years, there is a growing number of works on image reconstruction for QIS \cite{Chan16,gyongy2018single, chandramouli2018little, Ma_SIGGRAPH20,chi_chan_eccv2020}. Yang et al. \cite{yang2011bits} and Vogelsang et al. \cite{vogelsang2014hardware} showed how oversampling can be used to improve the dynamic range.  Another way to achieve a better dynamic range is to control the threshold dynamically, as proposed by Elgendy and Chan \cite{elgendy2018optimal}.The scheme we propose in this paper is to acquire multiple exposures and reconstruct the HDR image.

\begin{figure*}[t]
    \centering
    \includegraphics[width = 0.85 \linewidth]{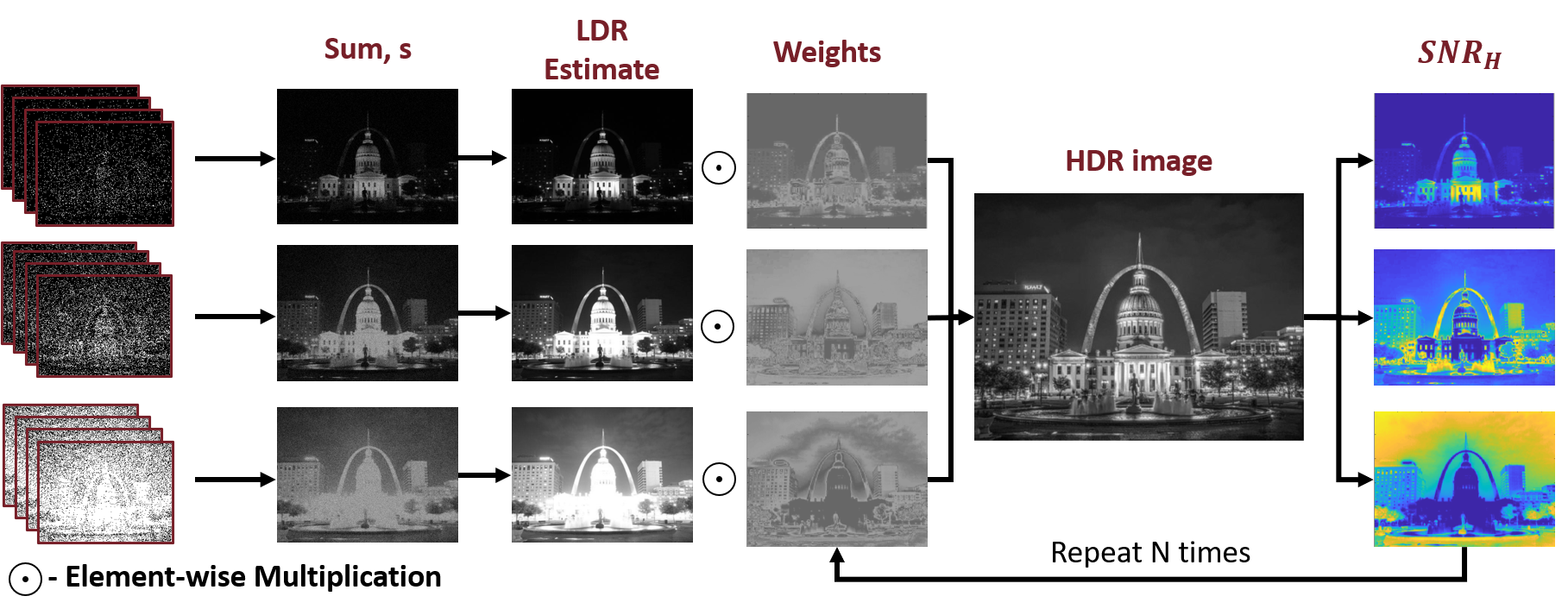}
    \caption{\label{fig:Reconstruction_pipeline}{\textbf{HDR Reconstruction Pipeline}}. The raw frames from QIS are first summed and denoised. Then the denoised images are linearly combined by giving weights to each image proportional to their $\text{SNR}_H$ iteratively. (See Section \ref{sec:HDR_rec})}
\end{figure*}

\subsection{HDR imaging using Quanta Image Sensors}

We present in this paper a QIS based HDR imaging principle, as outlined in \fref{fig:idea}. Within one CIS exposure, we use the QIS to over-sample the scene by taking multiple very short exposure frames of different integration periods. Depending on the full-well capacity and the scene dynamics, we can manually control the QIS to output 1-bit signals or few-bit signals by adjusting the analog-to-digital converter (ADC) \cite{fossum2015multi,fossum2013modeling}. The raw captures of the sensor are a stack of low bit-depth frames, some with longer exposures and some with shorter exposures. Because of the unique QIS statistics, we derive a customized image fusion algorithm to reconstruct an HDR image from the raw data.

There are several benefits of using QIS for HDR imaging compared to CIS. As we will theoretically derive in this paper, for the same total integration time, QIS offers a higher SNR compared to CIS. This, in turn, provides a wider dynamic range of the sensor. Besides, since QIS can capture multiple exposures within a short period, the overall acquisition time can be much shorter than that of CIS. For scenes containing moving objects, the short acquisition time of QIS is fundamentally more advantageous because it allows us to resolve the moving content while maintaining the dynamic range. Furthermore, by construction, the QIS has a much higher photon sensitivity than the CIS. Therefore, the QIS can handle low-light conditions much more effectively than the CIS.

\subsection{Single-photon avalanche diodes}
Single Photon Avalanche Diodes (SPAD) \cite{dutton2015spad,morimoto2020megapixel,bruschini2018monolithic} is an alternative technology to the CMOS-based QIS we use in this paper. SPAD are different from QIS in multiple ways. First, they amplify signals using avalanche multiplication. As such, SPAD require higher operating voltages(15-20V).  SPAD also have high dark current ($>10e^-/\text{pix}/\text{s}$), large pitch ($>5\mu$m), low fill-factor ($<70\%$), and low quantum efficiency ($<50\%$). In comparison, QIS do not use avalanche multiplication. It has smaller dark current($<0.1e^-/\text{pix}/\text{s}$), smaller pixel pitch ($1.1 \mu m$), higher fill factor ($>90 \%$), and higher quantum efficiency ($>70 \%$). SPAD have better frame-rate (upto 97k frames/sec) compared to QIS (1040 frames/sec), because of which SPAD are better suited for applications such as time-of-flight imaging, where resolving time stamps of the photon arrival is needed.

Using SPAD for HDR imaging has recently been demonstrated in several papers. Dutton et al. \cite{dutton2018high} showed a method to perform HDR imaging with active clock-driven SPAD image sensors using exposure bracketing. Ingle et al. \cite{ingle2019high} showed the suitability of active event-driven SPAD image sensors for passive HDR imaging. While SPAD have superior time resolving capability, QIS  offer higher spatial resolution. QIS can capture multi-bit frames, but SPAD can operate only in a single-bit mode. The QIS we use in this paper is clock-driven, so it is comparable to the SPAD used in \cite{dutton2018high}. However, \cite{dutton2018high} simply sums multiple frames to reconstruct an HDR image which is not optimal. The HDR reconstruction method we present here can be applied to the SPAD with some modifications of the parameters such as read noise, dark current, and bit-depth.

\subsection{Challenges and contributions}
QIS is arguably a young technology. While previous work of Yang et al. \cite{yang2011bits}, Fossum \cite{fossum2013modeling,fossum2015multi}, and Elgendy and Chan \cite{Chan16,elgendy2018optimal}  have laid the mathematical foundations of the sensor's sampling mechanisms and various image reconstruction algorithms, the sensor's performance for HDR imaging has never been systematically studied. The two biggest questions are: (i) to what extent the dynamic range can be offered by the QIS, and (ii) by what reconstruction algorithm we need to use for QIS. The goal of this paper is to fill the gap by outlining the theoretical performance limits of the sensors (Section \ref{sec:theory}), and propose an image fusion algorithm to merge the raw QIS frames into an HDR image (Section \ref{sec:HDR_rec}). By accomplishing this goal, we aim to provide a new HDR imaging approach.

The algorithmic contributions of this paper can be summarized in \fref{fig:Reconstruction_pipeline}. The input to the algorithm is a stack of images taken at different integration times. Assuming statistical independence of the measurements within the exposure stacks, previous work of Chan et al. \cite{Chan16} showed that the sum of the frames is a sufficient statistic of the scene. This allows us to create rough estimates of the low dynamic range (LDR) images. In this paper, we develop an iterative updating procedure to estimate the HDR image and predict the signal-to-noise ratio. The feedback loop continues until the algorithm converges. More specifically, we contribute to the literature of QIS and HDR imaging in the following ways:

\begin{enumerate}
\item We theoretically derive a closed-form expression for the signal-to-noise ratio (SNR) of QIS, with consideration of shot noise and read noise. We show that our theoretical prediction matches with the real data.
\item We use the SNR expression derived to compare the dynamic range of the QIS with CIS, from which we demonstrate the cost-benefit of QIS for high dynamic range imaging.
\item We develop a provably optimal HDR reconstruction algorithm that linearly combines the low dynamic range images, whose weights are obtained using the SNR values of each pixel at different integration times.
\end{enumerate}
The QIS camera we use in this paper is the \emph{PathFinder} camera developed by Gigajot Technology Inc. and Dartmouth College. Since the camera is still a prototype, we acknowledge its non-ideal optics and circuits.

\section{Background \label{section:model}}

\subsection{Imaging model of QIS}
The principle of QIS is to partition a pixel into many tiny cells called jots where each jot is a single-photon image sensor. Because of the small pixel pitch and the fast response, QIS can be regarded as an \emph{oversampling} device which oversamples the space and time. To understand how the sensors work, in this section we discuss their imaging model. Our model is more detailed than the previously studied models, e.g., \cite{Chan16,yang2011bits,elgendy2018optimal}, which focus mainly on the shot noise. We consider several other sources of noise. \tref{tab:notations} shows a list of notations used in this paper.

\begin{table}[h]
    \centering
    \caption{Notations used in this paper.}
    \begin{tabular}{cl}
    \hline
         Symbols &  Meanings  \\
         \hline
         d & Number of pixels \\
         $\vlambda$ & Photon flux from the scene\\
         T & Duty cycle of the camera \\
         $\Delta$ & Integration time \\
         $\vtheta$ & Mean number of photons arriving at the sensor\\
         $\mu_\text{dark}$ & Mean dark current\\
         $\mK$ & Number of photons detected by the sensor \\
         $\veta_\text{red}$ & Read noise from the image sensor\\
         $\sigma_\text{read}$ & Variance of the read noise\\
         $\mZ$ & Analog signal from the sensor\\
         $\mY$ & Digitized signal from the sensor\\
         $\mathcal{Y}$&Set of frames obtained at different integration times\\
         $\sigma_\text{H}$ & Exposure-referred noise\\
         $\text{SNR}_\text{H}$& Exposure-referred SNR \\
         $N$ & Total Number of frames collected\\
         $M$ & Number of different integration times\\
         $\Psi_q(.)$ & Incomplete Gamma function\\
         \hline
    \end{tabular}
    \label{tab:notations}
\end{table}

Let $\vlambda(t) = [\lambda_1(t),\ldots,\lambda_d(t)]^T \in \R^d$ be an $d$-dimensional vector field representing the photon flux with $d$ pixels located at time $t \in \R$. Let $T$ be the duty cycle when acquiring an image, and define the integration time as $[nT, nT+\Delta]$, for $n = 1,\ldots,N$. The mean number of electrons excited by the photons at the sensor is
\begin{equation}
    \vtheta[n] = \int_{nT}^{nT+\Delta} \vlambda(\tau) \; d\tau.
\end{equation}
The resulting vector field $\vtheta[n] = \big[\theta_1[n], \theta_2[n], \ldots, \theta_d[n]\big]^T$, where $n = 1,2,\ldots,N$, can be regarded as a video sequence of $d$ pixels and $N$ frames.

Given $\vtheta[n]$, we model the photon arrival as a Poisson process. Letting $\mK[n] = [K_1[n], \ldots, K_d[n]]^T$ be the number of photons detected by the sensor, we model $\mK[n]$ as
\begin{equation}
    \underset{\text{number of photons}}{\underbrace{\mK[n]}} \sim \qquad \underset{\text{Poisson process}}{\underbrace{\text{Poisson}\bigg( \vec{\theta}[n]+\mu_{\text{dark}} \bigg)}},
\end{equation}
where $\text{Poisson}(\cdot)$ denotes the Poisson distribution,
and $\mu_\text{dark}$ is the average dark current.

During the read out, we model the read noise as an i.i.d. Gaussian random variable, i.e., $\veta_{\text{read}}[n] \sim \calN(0,\sigma_\text{read}^2\mI)$. This leads to the  signal $\mZ[n] = \big[Z_1[n], \ldots, Z_d[n]\big]^T$ ,where
\begin{equation}
\mZ[n] = \underset{\text{number of photons}}{\underbrace{\mK[n]}} + \underset{\text{read noise}}{\underbrace{\veta_{\text{read}}[n]}}.
\end{equation}
Here, we denote the vector $\mZ[n]$ in upper case to emphasize that it is a random vector. The distribution of $\mZ[n]$ is Poisson-Gaussian. Specifically, letting $\vartheta_i = \theta_i[n] + \mu_{\text{dark}}$ be the Poisson mean, the probability of $Z_i[n] = z$ is given by
\begin{equation}
\Pb[Z_i[n] = z] = \sum_{\ell=0}^{\infty}\bigg({\frac{\vartheta_i^\ell}{\ell!}e^{-\vartheta_i} \cdot \frac{1}{\sqrt{2\pi \sigma_\text{read}^2}}e^{-\frac{(z-\ell)^2}{2\sigma_\text{read}^2}}}\bigg).
\label{eq: P[Z_i = z]}
\end{equation}

Finally, to generate a digital signal, we convert the real number $\mZ[n]$ using an analog-to-digital converter (ADC). Depending on the number of bits used by the ADC, we can operate the QIS in a single-bit mode or a multi-bit mode. For \textbf{single-bit}, the ADC binarizes the signal as
\begin{equation}
    Y_i[n] \bydef \text{ADC}\bigg(Z_i[n]\bigg)
    =
    \begin{cases}
    1, &\quad\text{if}\;\; Z_i[n] \ge L,\\
    0, &\quad\text{if}\;\; Z_i[n] < L.
    \end{cases}
\end{equation}
For \textbf{multi-bit}, the ADC returns
\begin{equation}
    Y_i[n] \bydef \text{ADC}\bigg(Z_i[n]\bigg)
    =
    \begin{cases}
    L, &\round{Z_i[n]} \ge L,\\
    \round{Z_i[n]}, &\round{Z_i[n]} < L,
    \end{cases}
\end{equation}
where $\round{.}$ is function that rounds off the real valued numbers to the nearest integer. The final raw QIS signal generated is therefore a sequence of vectors
\begin{equation}
\mathcal{Y} = \bigg\{
\underset{\text{Exposure 1}}{\underbrace{\mY[1],\ldots,\mY[m]}} \;\; \ldots \;\;\underset{\text{Exposure M}}{\underbrace{\mY[n],\ldots,\mY[N]}} \bigg\},
\label{eq: calY}
\end{equation}
where each $\mY[n] = [Y_1[n],\ldots,Y_d[n]]^T$. Our goal is to reconstruct $\vtheta[n]$ from $\mathcal{Y}$. For repeated measurements of the same scene, we will have multiple realizations of $\mathcal{Y}$.

\subsection{Model validation with real data}
The model presented above is more comprehensive than the previous models in \cite{Chan16} and \cite{yang2011bits}. To validate our model we compare it with the real data collected by a prototype QIS camera. The experiment was first reported in \cite{ma2015quanta}, and we repeat the experiment again in this paper for completeness.

\begin{figure}[ht]
    \centering
    \vspace{-2ex}
    \includegraphics[width=\linewidth]{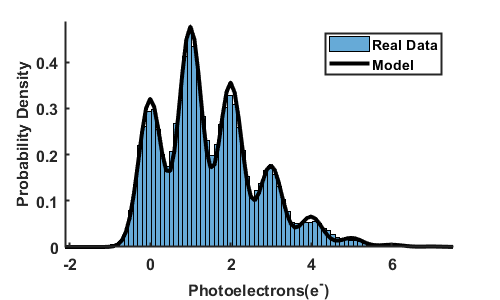}
    \vspace{-4ex}
    \caption{\textbf{Validation of the Model.} First reported in \cite{ma2015quanta}. We compute the photon counting histogram of a real QIS sensor and compare it with our theoretical model. Note the similarity between the two. }
    \label{fig:PCH}
\end{figure}

In this experiment, we use a uniform illumination so that $\vlambda[n]$ is a constant. A total of 50,000 repeated measurements from a single pixel is used to construct a photon counting histogram as shown in \fref{fig:PCH}. Each measurement has an integration time of $50 \mu$s, The average photon count is 1.48 photons per pixel (ppp). The ADC uses a bit-depth of 14 bits. The least significant bit is 0.05e$^{-}$. Because the ADC uses 14 bits, the resulting histogram is close to a continuum.

To plot the theoretical model, we assume that the read noise level is 0.25$e^-$. The dark current is assumed to be 0.0068$e^{-}$ per second \cite{Ma:17}. Now, once we know the scene intensity $\lambda$, we can plot the theoretical probability density function. $\lambda$ is chosen such that the mean squared error between the histogram and the theoretical curve is minimized. Since the integration time is only 50$\mu$s, we can safely neglect the dark current. Putting these together we obtain the black curve as shown in \fref{fig:PCH}. As we can see, the theoretical model fits the real data well.

\section{Theoretical Analysis of QIS Dynamic Range} \label{sec:theory}
\begin{figure}[t]
    \centering
    \includegraphics[width = 0.9\linewidth]{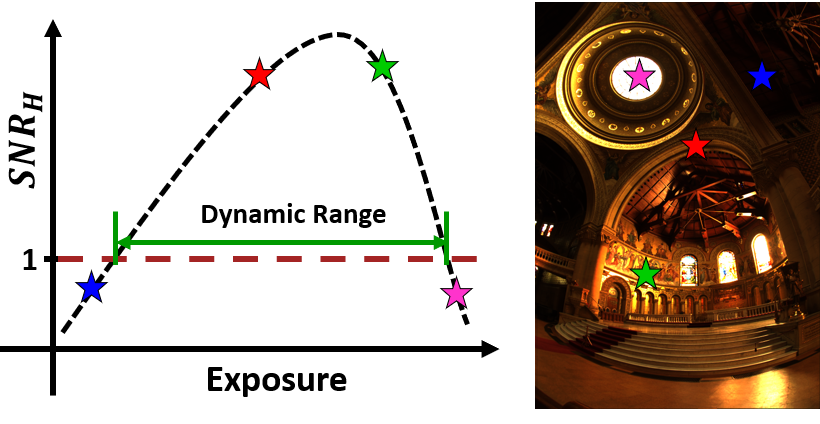}
    \caption{\textbf{What is Dynamic Range?} Dynamic range is the range of exposure that can be detected by the sensor. We define it as the range of exposure for which the SNR is greater than 1. If a particular part of a scene has lower exposure than this range, then it will appear black. Similarly, excessive exposure may make a pixel appear saturated. The image on the right is taken from \cite{debevec2008recovering}.  }
    \label{fig:dynamic_demo}
\end{figure}

In this section, we theoretically derive the dynamic range of a QIS. \fref{fig:dynamic_demo} shows the meaning of the dynamic range. The dynamic range is the range of the exposure such that the signal-to-noise ratio (SNR) is above a certain threshold. Therefore, to analyze the dynamic range of QIS, we need to first define the SNR and then determine how the SNR changes when different sensor parameters change.

\subsection{Signal to noise ratio (SNR)}
To simplify our notation, we will focus on the $i$-th pixel $Y[n] = Y_i[n]$, for $n = 1,\ldots,N$. Since the sensor response of QIS is a nonlinear, we follow \cite{fossum2013modeling} and \cite{elgendy2018optimal} by considering the exposure-referred SNR.

\begin{definition}[Exposure-referred SNR]
Let $Y[n]$ be a pixel having an average illumination of $\theta$ photons. The exposure-referred signal-to-noise ratio $\text{SNR}_{\text{H}}$ is defined as
\begin{align}
\text{SNR}_{\text{H}}(Y[n])
&= \frac{\text{signal}}{\text{expo-ref noise}} \notag\\
&\bydef \frac{\theta}{\sigma_H} = \frac{\theta}{\sigma_Y} \frac{d \mu_Y}{d\theta}
\label{eq: SNR}
\end{align}
where $\mu_Y = \E[Y[n]]$ and $\sigma_Y = \sqrt{\Var[Y[n]]}$ are the mean and standard deviation of the signal $Y[n]$, respectively, and $\sigma_H$ is the exposure referred noise defined as $\sigma_H \bydef \sigma_Y \frac{d\theta}{d\mu_Y}$.
\end{definition}

We choose to use $\text{SNR}_\text{H}$ because, for any random quantity $Y$, the typical output-referred signal-to-noise ratio is $\E[Y]/\sqrt{\Var[Y]}$. For nonlinear sensors such as QIS, output-referred signal-to-noise ratio explodes as the truncated Poisson random variable $Y$ will have an infinite SNR when exposure increases. Exposure-referred SNR ensures that very long exposure will have negative effects on the sensor because pixels could be saturated.

When summing a total of $N$ independent frames that are acquired using the same integration time, we define
\begin{equation}
    S = \sum_{n=1}^N Y[n].
\end{equation}
It is easy to show that the expectation and variance are
\begin{align*}
\E[S] = N\mu_Y, \qquad \mbox{and} \qquad \Var[S] = N \sigma_Y^2.
\end{align*}
In this case, the SNR becomes
\begin{equation}\label{eqn:SNRH_defn}
\text{SNR}_{\text{H}}(S) = \sqrt{N}\frac{\theta}{\sigma_Y} \frac{d \mu_Y}{d\theta}.
\end{equation}
We now present the main theoretical result. Theorem~\ref{thm:mub_sigb} shows the detailed quantities of the SNR, namely $\mu_Y$, $\sigma_Y$ and $d\mu_Y/d\theta$ for single-bit and multi-bit truncated Poisson random variables.

\begin{theorem}\label{thm:mub_sigb}
Let $Y[n]$ be a multi-bit QIS measurement, i.e.,
$$
\text{ADC}\big(Z_i[n]\big) =
\begin{cases}
L       &\qquad\mbox{if}\; \round{Z_i[n]} \ge L,\\
Z_i[n]  &\qquad \mbox{if}\; \round{Z_i[n]} < L.
\end{cases}
$$
Then it holds that
\begin{align}
\mu_{Y} &= \theta(\Psi_{L-1}(\theta )) + L(1 - \Psi_{L}(\theta)) + \Delta_\mu(\theta), \label{eq: E[Y]}\\
\sigma_{Y}^2 &=  L^2 - \sum\limits_{q=0}^{L-1} ((2q+1) \Psi_{q+1}(\theta)) + \Delta_{\sigma^2}(\theta) -  \mu_{Y}^2,
\label{eq: Var[Y]}
\end{align}
where $\Psi_q(\theta) = \sum\limits_{k=0}^{q-1}\frac{\theta^k e^{-\theta}}{k!}$ is the incomplete gamma function \cite{abramowitz1948handbook}, and $\theta$ is the underlying Poisson mean defined in \eref{eq: P[Z_i = z]} \footnote{For notation simplicity we ignore the dark current.}. The quantities $\Delta_\mu(\theta)$ and $\Delta_{\sigma^2}(\theta)$ are respectively
\begin{align}
\nonumber \Delta_\mu(\theta) = &\sum\limits_{k=-\infty}^\infty p_k \bigg(  \sum\limits_{q = [k]_+}^{L-1} \bigg( \frac{e^{-\theta} \theta^{q-k}}{(q-k) !} - \frac{e^{-\theta} \theta^q}{q !}\bigg) q \\
& + L(\Psi_L(\theta) - \Psi_{[L-k]_+}(\theta))\bigg)\\
\nonumber \Delta_{\sigma^2}(\theta) = &\sum\limits_{k=-\infty}^\infty p_k \bigg(   \sum\limits_{q = [k]_+}^{L-1} \bigg( \frac{e^{-\theta} \theta^{q-k}}{(q-k) !} - \frac{e^{-\theta} \theta^q}{q !}\bigg) q^2  \\
&+ L^2(\Psi_L(\theta) - \Psi_{[L-k]_+}(\theta))\bigg),
\label{eq: dmu/dtheta}
\end{align}
where $[\cdot]_+ = \max(\cdot,0)$ returns the positive value, and
\begin{equation}
    p_k = \int_{k-0.5}^{k+0.5} \frac{1}{\sqrt{2\pi \sigma_\text{read}^2} }e^{-\frac{x^2}{2\sigma_\text{read}^2}} dx
\end{equation}
is the error probability due to read noise. The derivative $d\mu_Y/d\theta$ is
\begin{align}
\nonumber &\frac{\partial \mu_Y}{\partial \theta} = \Psi_{L-1}(\theta) - \theta\frac{e^{-\theta} \theta^{[L-2]_+}}{[L-2]_+!} +  L \frac{e^{-\theta} \theta^{[L-1]_+}}{[L-1]_+!} \\
\nonumber &+\sum\limits_{k=-\infty}^\infty p_k \bigg(  \sum\limits_{q = [k]_+}^{L-1} \bigg( - \frac{e^{-\theta} \theta^{q-k}}{(q-k) !}  + \frac{e^{-\theta} \theta^q}{q !}\bigg) q \\
\nonumber &+ \sum\limits_{q = [k]_+}^{L-2} \bigg( \frac{e^{-\theta} \theta^{q-k}}{(q-k) !}  - \frac{e^{-\theta} \theta^q}{q !}\bigg) (q+1) \\
 &+ L\big( -\frac{\theta^(L-1) e^{-\theta}}{(L-1)!} + \frac{\theta^([L-k-1]_+) e^{-\theta}}{[L-k-1]_+!}\big) \bigg).
\end{align}
\end{theorem}
\begin{proof}
See supplementary document \cite{gnanasambandam2020hdr}.
\end{proof}

The expressions offered by Theorem~\ref{thm:mub_sigb} are dense. However, they are also \emph{exact}. This is a generalization of the previous work by Gnanasambandam et al. \cite{gnanasambandamhigh} which showed a special case of our theorem without considering the read noise.

\subsection{Understanding Theorem~\ref{thm:mub_sigb}}
To help readers gain some insights about Theorem~\ref{thm:mub_sigb}, we discuss a few aspects of the theorem.

\begin{figure}[ht]
    \centering
    \includegraphics[width=\linewidth]{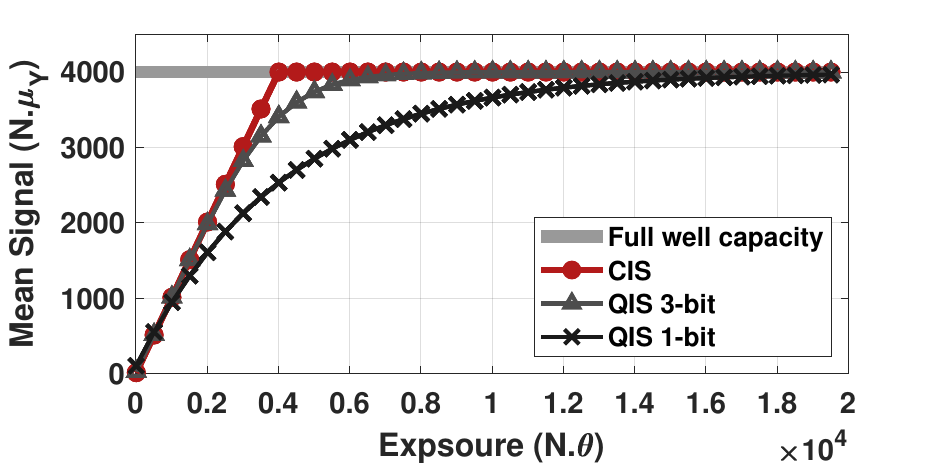}
    \vspace{-4ex}
    \caption{\textbf{Sensor Response.} The mean signal $\E[Y[n]]$ of a CIS and a QIS, as a function of the exposure $\theta$.}
    \label{fig:mu_compare}
\end{figure}

\textbf{Sensor response of QIS and CIS}. We first discuss the measured signal of a QIS compared to a CIS. As we discussed in the background section, a QIS pixel has a nonlinear response driven by the Poisson-Gaussian statistics, subject to an ADC. Theorem~\ref{thm:mub_sigb} shows that this signal $Y[n]$, although being random, has a mean $\E[Y[n]]$ given by \eref{eq: E[Y]}:
\begin{equation}
\text{QIS:} \quad \E[Y[n]] =
    \theta(\Psi_{L-1}(\theta )) + L(1 - \Psi_{L}(\theta)) + \Delta_\mu(\theta).
\end{equation}
We can plot this mean signal $\E[Y[n]]$ as a function of $\theta$. The result is shown in \fref{fig:mu_compare}. Depending on the bit-depth of the sensor, the mean $\E[Y[n]]$ demonstrates a soft saturation as the exposure $\theta$ approaches the full well capacity.

In the same figure, we show the sensor response of a CIS.  CIS has a linear response with respect to the exposure $\theta$.
\begin{equation}
\text{CIS:} \quad    \E[Y[n]] =
    \begin{cases}
    \theta, &\qquad \theta \le L,\\
    L, &\qquad \theta > L,
    \end{cases}
    \hspace{2.5cm}
\end{equation}
where $L$ is the full well capacity of the sensor. \fref{fig:mu_compare} shows that because of the linear response, the signal of CIS rises linearly and then saturates once it hits the full well capacity.

\begin{figure*}[ht]
\centering
\begin{tabular}{cc}
\includegraphics[width=0.46\linewidth]{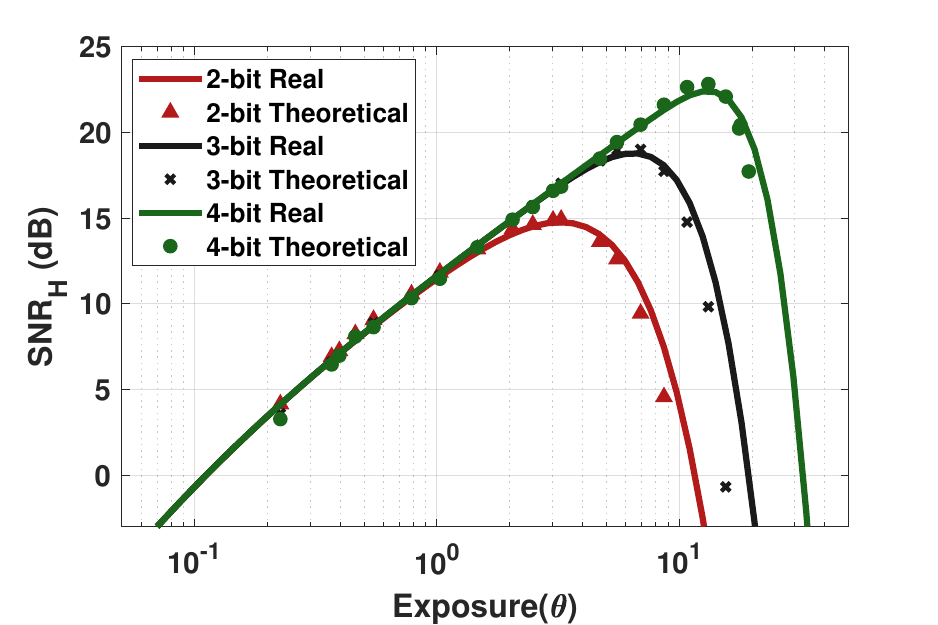}&
\includegraphics[width=0.46\linewidth]{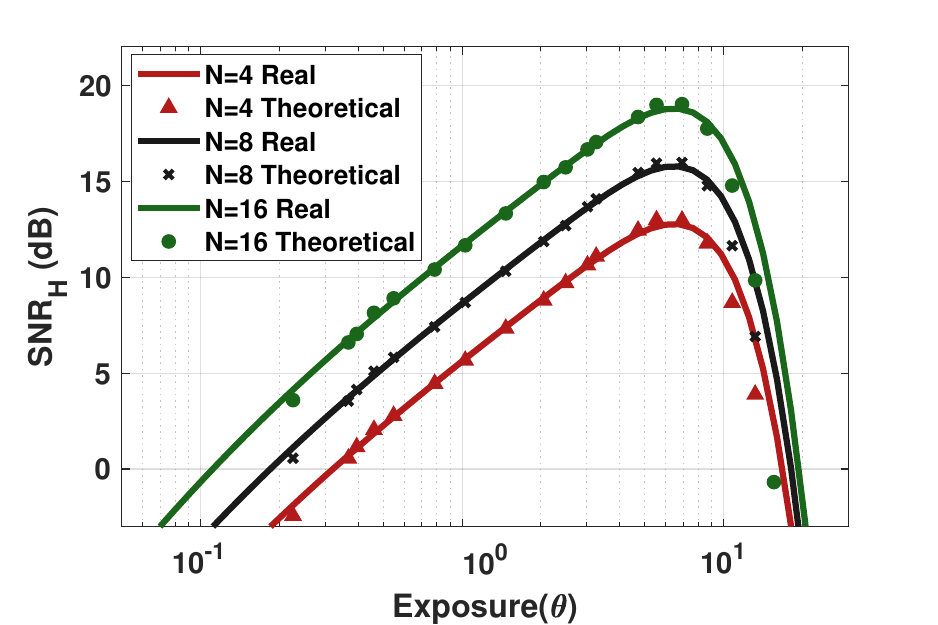}\\
(a) SNR for different bit-depth & (b) SNR for different $N$
\end{tabular}
\caption{\textbf{Theoretical vs Real $\SNRH$} (a) at different bit level. $T = 16$ frames were used to obtain all the curves. (b) with different number of frames. 4-bit data is used for obtaining all the curves. These two figures validate the correctness of Theorem~\ref{thm:mub_sigb}. Notice the soft saturation of the QIS, where the SNR drops smoothly over a range of exposure levels, whereas for a CIS, there will be a sudden drop in SNR as the exposure level reaches the full-well capacity.}
\label{fig:SNR_validation}
\end{figure*}

\textbf{How SNR changes w.r.t. bit depth}. Theorem~\ref{thm:mub_sigb} shows that $\text{SNR}_H$ changes with the bit-depth. This effect is shown in \fref{fig:SNR_validation} (a). Here, we plot both the real measured QIS data and the theoretically predicted curves using different bit-depths. All curves are plotted using $N = 16$ frames. The results indicate two things: (i) The theoretical prediction matches very well the actual QIS measurement. (ii) $\text{SNR}_H$ increases as the bit-depth increases. The latter happens because the term $\frac{d \mu_Y}{d \theta}$ drops faster for smaller bit-depth and slower for larger bit-depth. However, there is a trade-off between bit-depth and speed --- 2-bit data take a much shorter time to acquire than 4-bit data.

\textbf{How SNR changes w.r.t. the number of frames}. Theorem~\ref{thm:mub_sigb} provides information about the number of frames. We plot the theoretically predicted curves on top of a set of real 4-bit QIS data points. Here, we use a fixed integration time $\Delta$ for all choices of $N$, i.e., the total exposure for $N = 4$ is $4 \times \Delta$. \fref{fig:SNR_validation} (b) shows the results. As $N$ increases, we observe that $\text{SNR}_H$ also increases as predicted by the $\sqrt{N}$ term in the Theorem.

\begin{figure*}[ht]
\centering
\begin{tabular}{cc}
\includegraphics[width=0.45\linewidth]{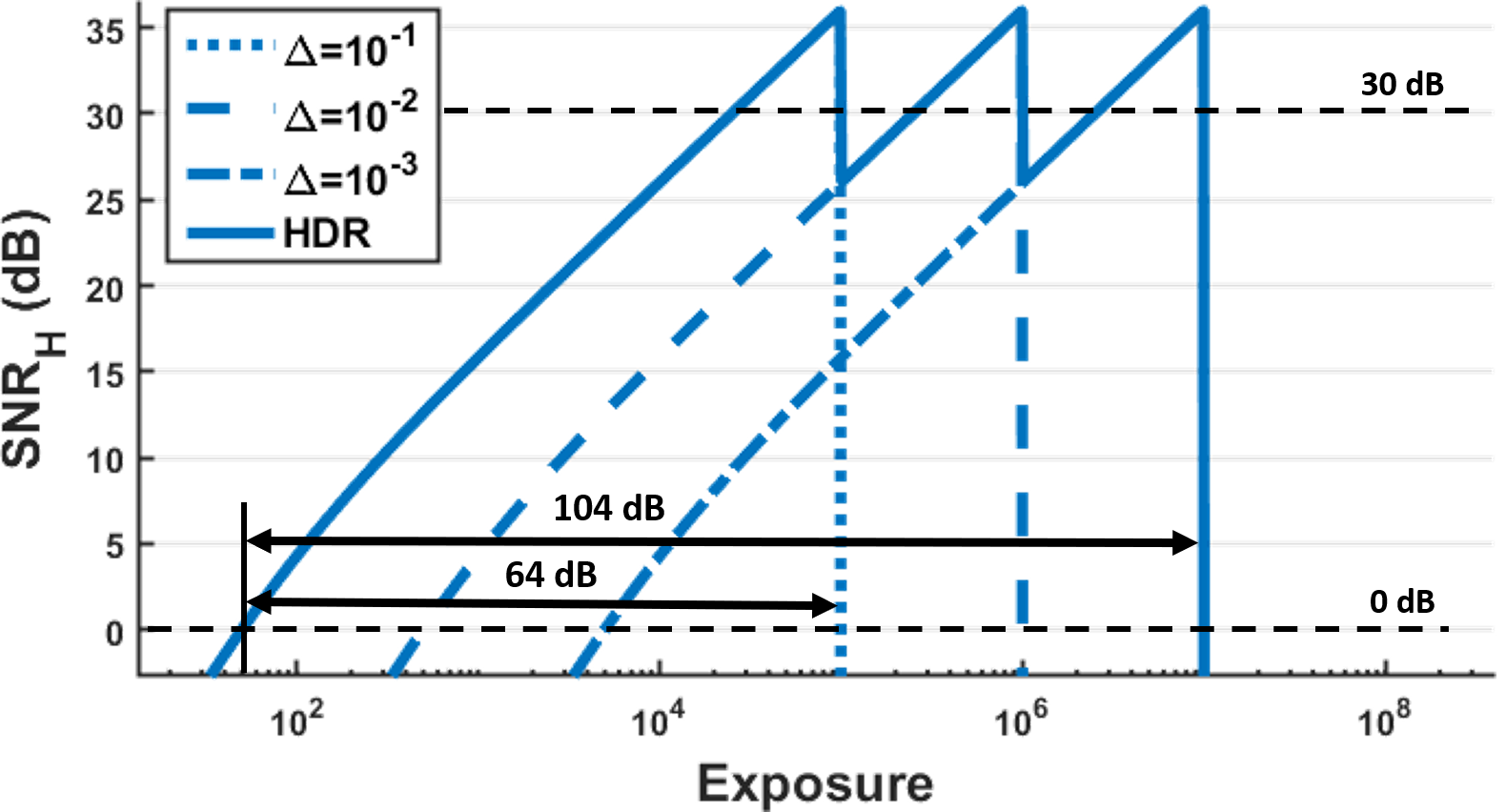}&
\includegraphics[width=0.45\linewidth]{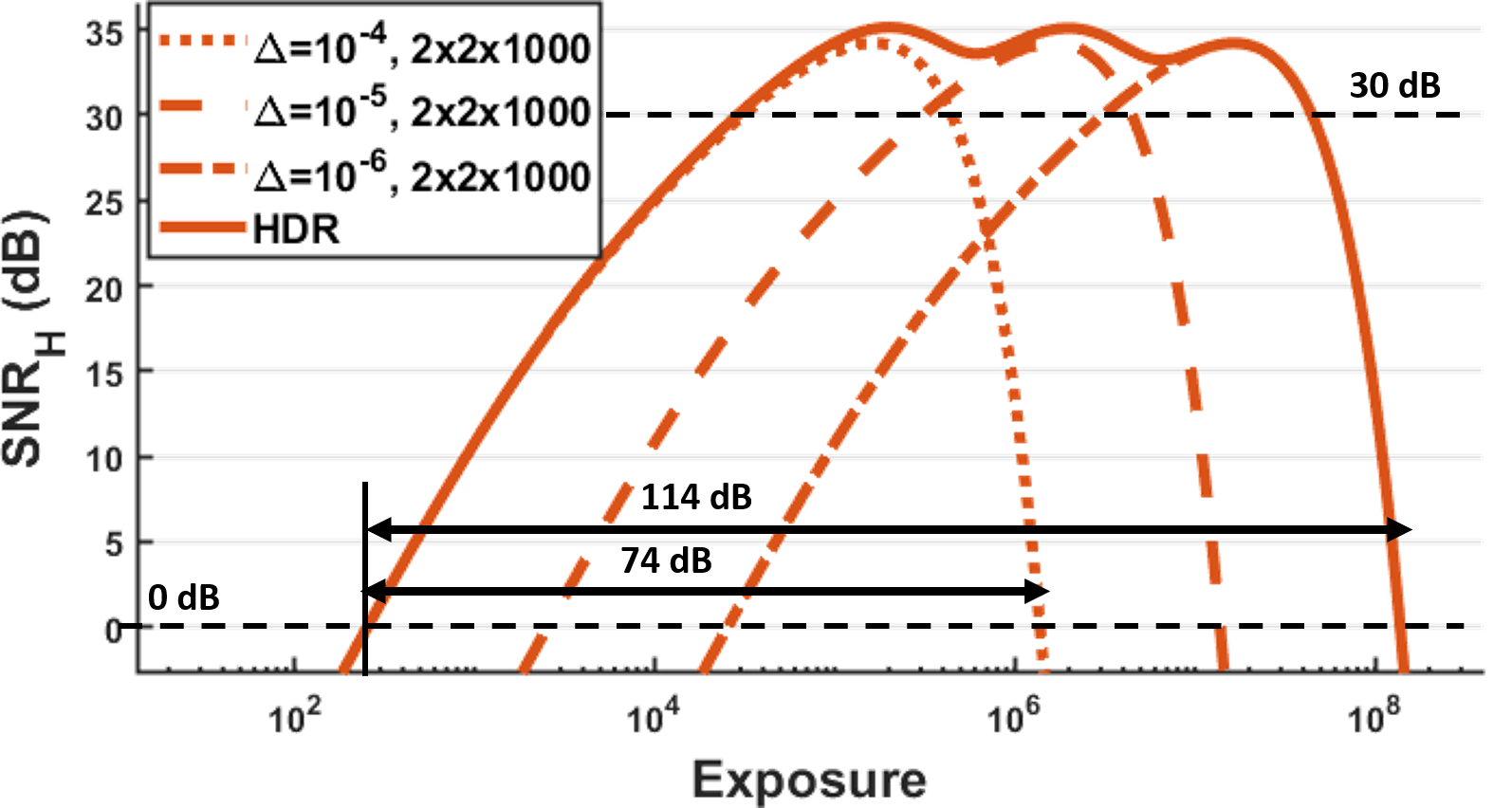}\\
(a) CIS & (b) 1-bit QIS
\end{tabular}
\caption{\textbf{Comparison of Exposure Referred Signal-to-Noise Ratio ($\text{SNR}_H$) for CIS and QIS}. CIS is assumed to have a full well capacity of 4000 electrons. QIS is assumed to use a spatial oversampling of $2\times 2$. The number of frames at each integration time is $T = 1000$ for single-bit QIS. The oversampling is chosen such that the total signal obtained by both the CIS and QIS is the same. Notice that the QIS has a larger dynamic range than CIS for each exposure, and has a more consistent SNR over the entire range when the low dynamic range images are combined to get a single high dynamic range image. }
\label{fig:SNR_CISvsQIS}
\end{figure*}

\subsection{Dynamic range of QIS and CIS}
\label{sec:expt}

Using Theorem~\ref{thm:mub_sigb}, we compare the dynamic range of a QIS and a CIS. Recall \fref{fig:dynamic_demo}, the dynamic range of a sensor is defined as the range of exposure such that the SNR is above unity. Our goal here is to use the theoretical curves to predict how much dynamic range can be offered by each sensor.

Considering a typical setup of a CIS where the full-well capacity is $L=4000$ $e^-$ and the read-noise is $\sigma_\text{read} = 2e^-$. We assume that the CIS uses three exposures to capture the image. For QIS, we operate in an oversampling regime by taking multiple short exposures of equal length. The number of frames is configured such that the total duration of the acquisition is the same as a CIS. Afterward, we merge the short-exposure frames to generate an HDR image using the algorithm to be described in Section~\ref{sec:HDR_rec}.

\begin{figure*}[!]
\centering
\begin{tabular}{ccc}
\hspace{-2.0ex}\includegraphics[width=0.32\linewidth]{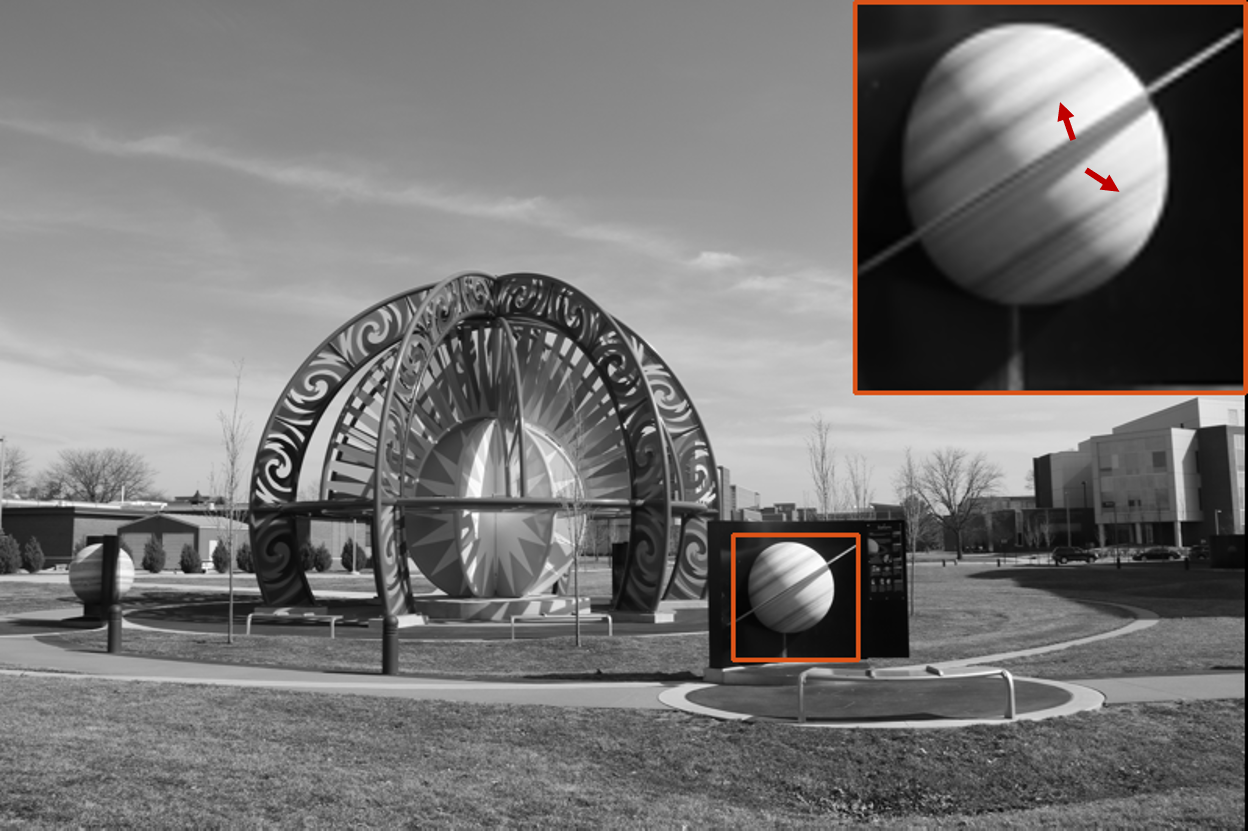}&
\hspace{-2.0ex}\includegraphics[width=0.32\linewidth]{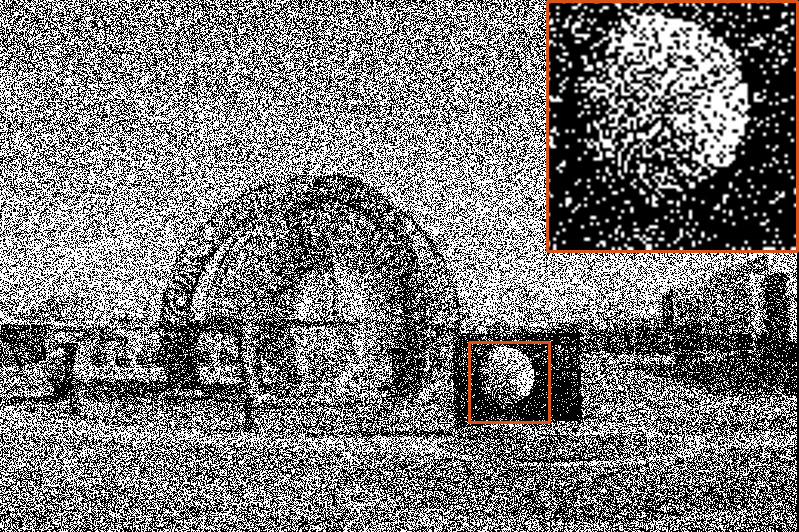}&
\hspace{-2.0ex}\includegraphics[width=0.32\linewidth]{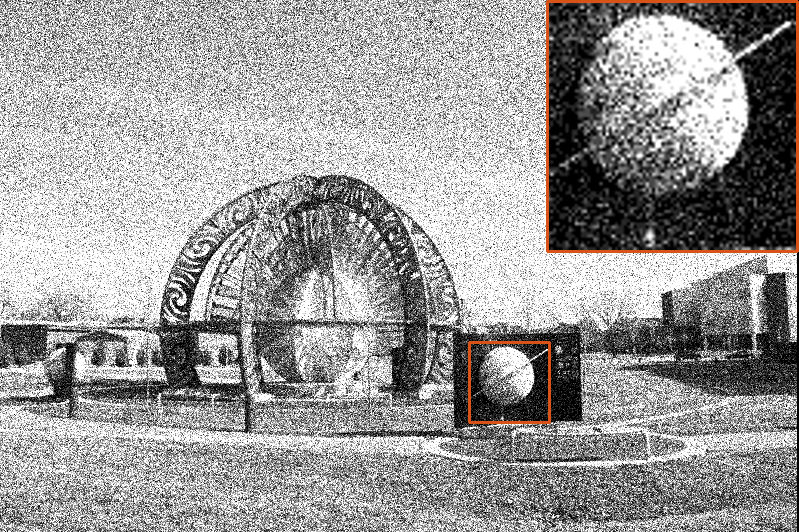}\\
(a) Ground Truth & (b) raw 1 frame (1-bit) & (c) raw 1 frame (3-bit) \\
\hspace{-2.0ex}\includegraphics[width=0.32\linewidth]{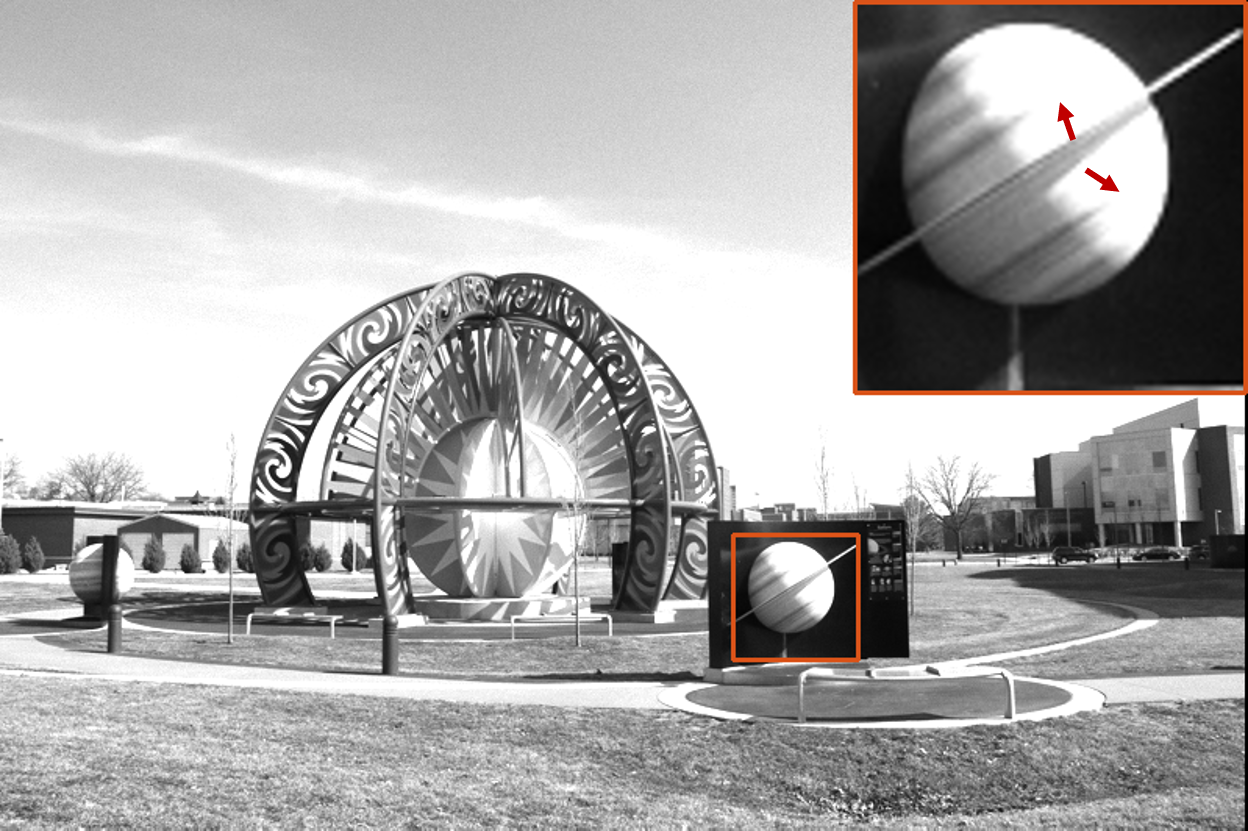} &
\hspace{-2.0ex}\includegraphics[width=0.32\linewidth]{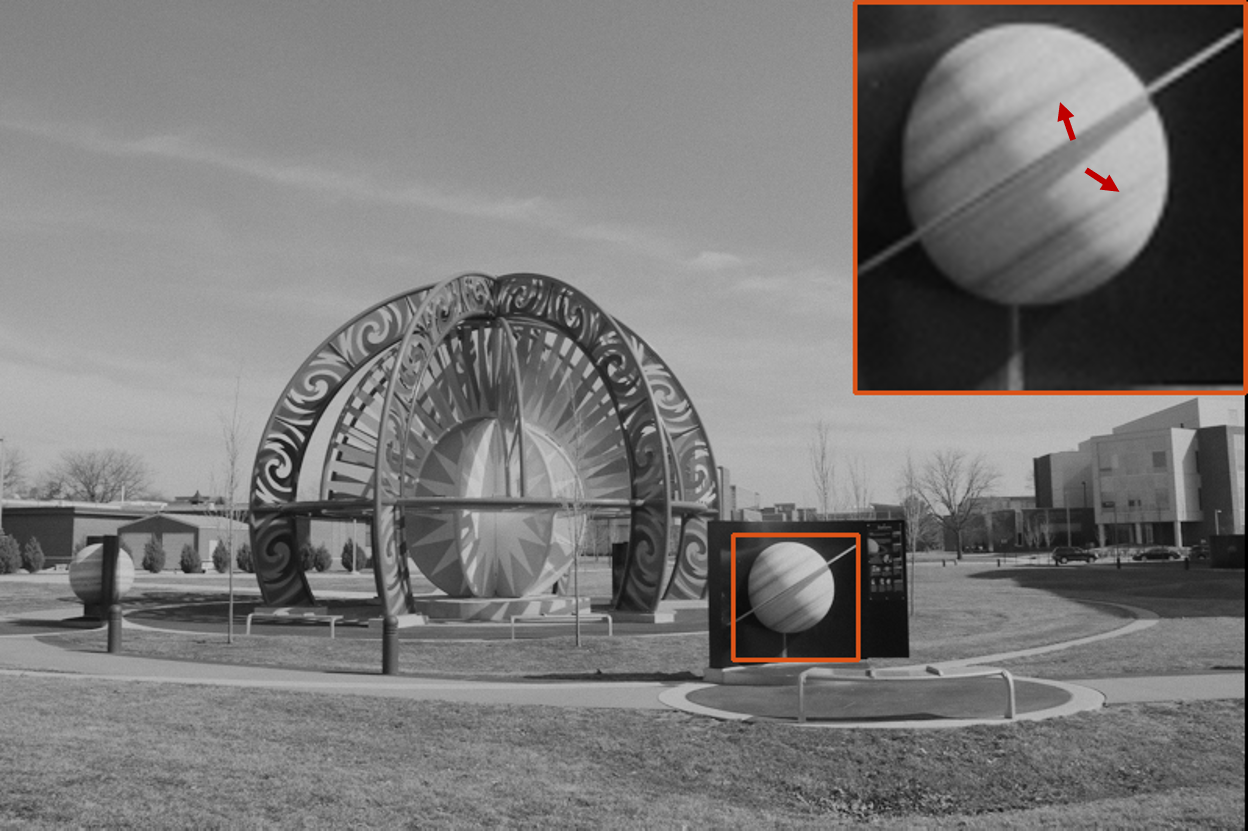}&
\hspace{-2.0ex}\includegraphics[width=0.32\linewidth]{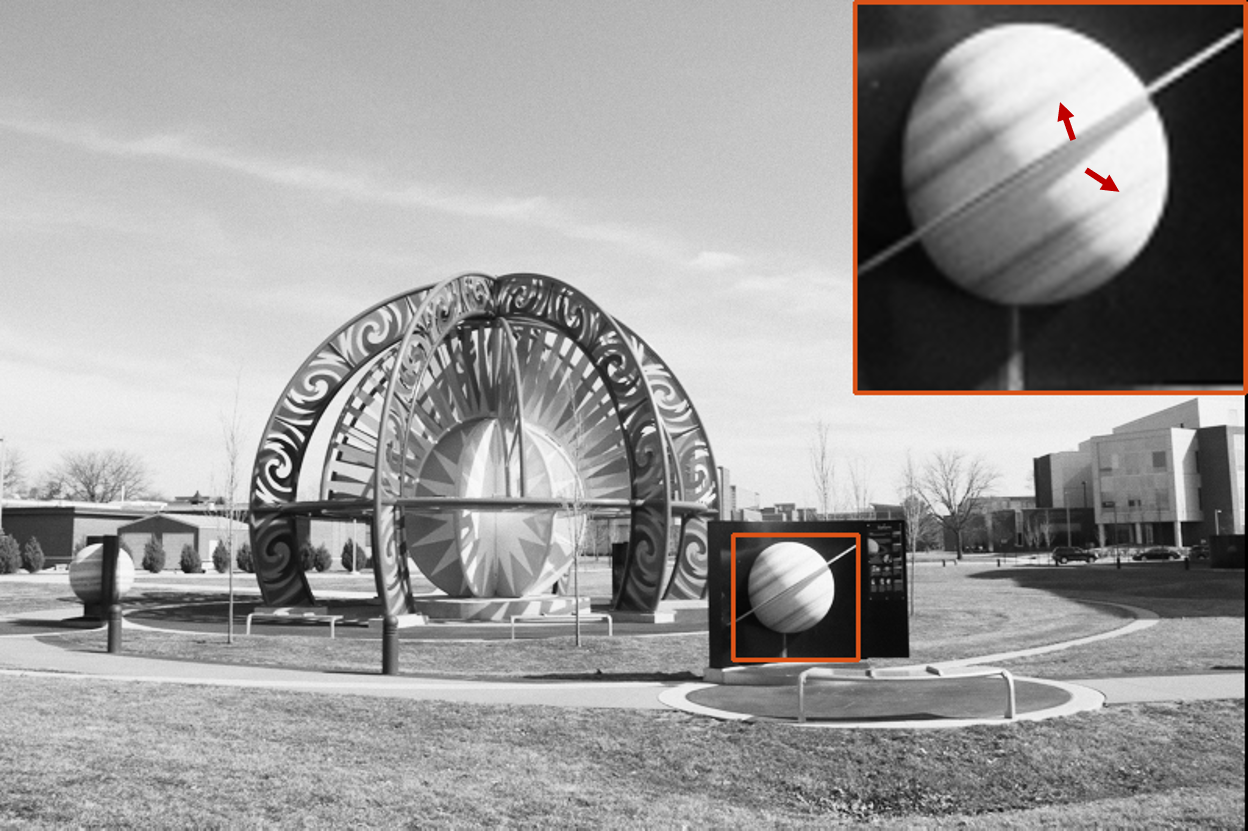}\\
(d) CIS & (e) sum of 4000 frames (1-bit) & (f) sum 571 frames 3-bit)
\end{tabular}
\caption{ \textbf{Dynamic range of QIS and CIS.}  The image is simulated in such a way that the maximum illumination of the image is $6 \times 10^6$ photons per pixel per second. CIS can count upto 4000 electrons, single bit QIS - 1 electron and 3 bit QIS - 7 electrons. The exposure times are: CIS - 1ms, single bit QIS - $0.25 \mu s$, and 3 bit QIS - $1.75 \mu s$. We use 1 CIS frame, 4000 frames for single bit QIS and 571 frames for 3 bit QIS. In the red arrowed, we observe that CIS is saturated whereas QIS still shows the signal.}
\label{fig:DR_cpmpare}
\end{figure*}

\begin{figure}[t]
\centering
\begin{tabular}{cccc}
\hspace{-2.0ex}\rotatebox{90}{\ \ \ \ \ \ \ {\fontsize{7}{7} \textbf{$\Delta=0.2, N=100$}}} &
\hspace{-2.0ex}\includegraphics[width=0.30\linewidth]{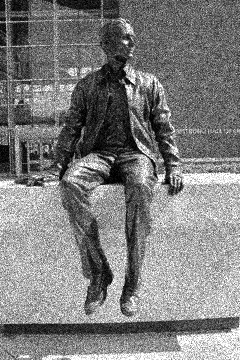}&
\hspace{-2.0ex}\includegraphics[width=0.30\linewidth]{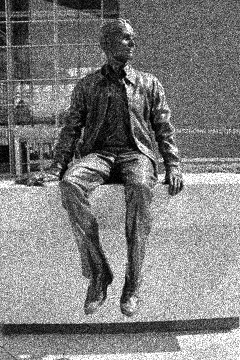}&
\hspace{-2.0ex}\includegraphics[width=0.30\linewidth]{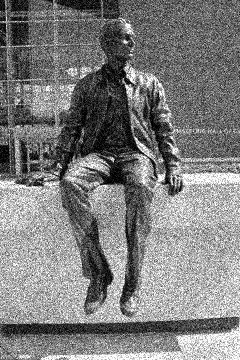}\\

\hspace{-2.0ex}\rotatebox{90}{\ \ \ \ \ \ {\fontsize{7}{7} \textbf{$\Delta=0.02, N=1000$}}} &
\hspace{-2.0ex}\includegraphics[width=0.30\linewidth]{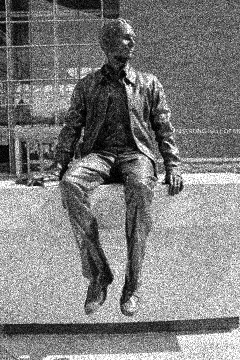}&
\hspace{-2.0ex}\includegraphics[width=0.30\linewidth]{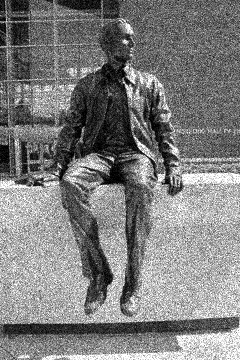}&
\hspace{-2.0ex}\includegraphics[width=0.30\linewidth]{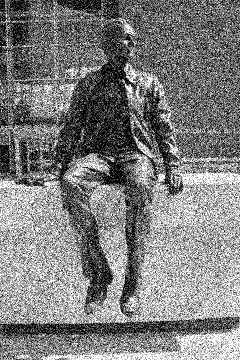}\\
 &$\sigma_{\text{read}} = 0$ & $\sigma_{\text{read}} = 0.15$ & $\sigma_{\text{read}} = 0.25$\\
\end{tabular}
\caption{\textbf{Accumulation of noise.} The sub-figures show the  sum of $N$ simulated QIS frames  when using different $N$ and different integration time $\Delta$, such that the total integration time $N\Delta$ is same for all the cases considered. Because of the finite read noise, by summing more frames we accumulate error. This leads to a trade-off between the number of frames and the SNR, when the total integration time is fixed. }
\label{fig:noise_accu}
\end{figure}

\fref{fig:SNR_CISvsQIS} shows the theoretically predicted curves for CIS and 1-bit QIS. For CIS, we show three different integration time $\Delta  = 10^{-1}$sec, $10^{-2}$sec and $10^{-3}$sec. The HDR image formed by a CIS is the sum of the three exposures. QIS with an integration time of $10^{-4}$ uses an oversampling of $2\times2\times 1000$. This means a spatial oversampling of a $2\times 2$ bin, and 1000 frames of 1-bit measurements. The HDR versions of the QIS data are obtained used the reconstruction method described in section \ref{sec:HDR_rec}.

As we can observe from \fref{fig:SNR_CISvsQIS}, the dynamic range of a QIS using just one integration time is 74dB (1-bit), which is already substantially larger than the 64dB of a CIS. After reconstructing the HDR image by merging multiple integration times, the resulting dynamic range offered by a QIS is also higher than that of a CIS. Also,  \fref{fig:SNR_CISvsQIS} shows that the SNR of a combined QIS image never drops below 30dB. This is a big contrast to CIS which has a sudden drop once the exposure exceeds the full-well capacity. \fref{fig:DR_cpmpare} illustrates the visual comparison between a CIS and a QIS. Notice that for the same amount of photons, the QIS offers better details than a CIS.

If we look at the low-light ends of \fref{fig:SNR_CISvsQIS}, we observe that CIS is performing better than a QIS. This phenomenon is the result of accumulating read noise from adding multiple frames. Since, every readout of a QIS frame has a fixed amount of read noise, the more readouts we do the more read noise we accumulate. In \fref{fig:noise_accu} we demonstrate this problem. Assuming a read noise level of $\sigma_\text{read} = 0.25$, and an integration time of $\Delta = 0.2$sec (or $\Delta = 0.02$sec), we plot the sum of $N$ frames of 3-bit frames. As the number of frames $N$ increases, with the total integration time fixed, the image becomes noisier when $\sigma_\text{read} = 0.25$. This is not visible, when $\sigma_\text{read} = 0$ or $\sigma_\text{read} = 0.15$, because the gaussian read noise is not strong enough to cause any issues with multiple read-outs.

\subsection{SNR vs dynamic range trade-off}
\begin{figure}[t]
    \centering
    \includegraphics[width = \linewidth]{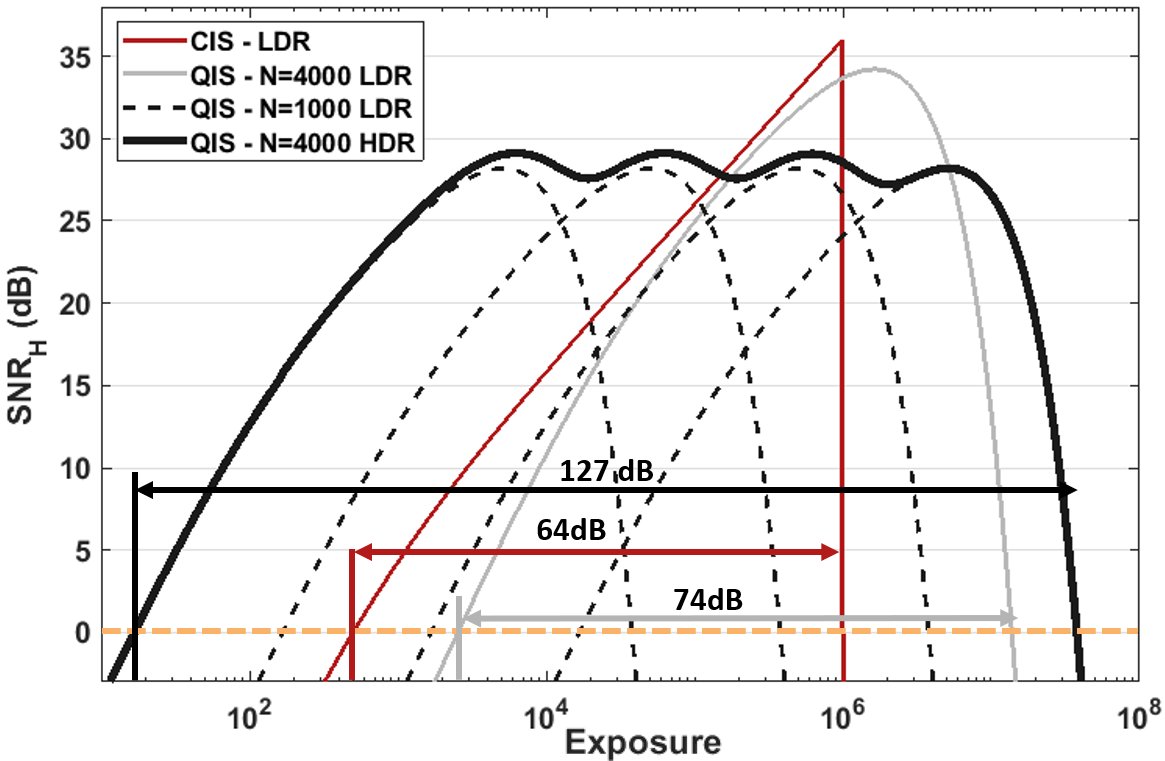}
    \vspace{-2.0ex}
    \caption{\textbf{SNR vs. Dynamic Range Tradeoff.} QIS offers a unique trade-off where we can choose a setting based on whether we want an image with very high SNR or large dynamic range. In this figure, we can see that QIS can operate under an LDR regime with comparable SNR to a CIS or HDR regime, where the dynamic range is significantly higher.}
    \label{fig:tradeoff}
\end{figure}

QIS offers a trade-off between the peak SNR and the dynamic range. \fref{fig:tradeoff} shows four sets of curves: (i) a CIS running in LDR mode. The dynamic range is 64dB. (ii) A QIS operating in LDR mode, with $N = 4000$ frames which is equivalent to a CIS exposure. The dynamic range is 74dB, but the peak SNR is slightly lower than that of CIS. (iii) A QIS operating in an LDR mode, with $N = 1000$ frames. This is a much weaker signal than a CIS. (iv) A QIS operating in the way as the previous case, but this time we merge four different LDR images to create an HDR image. We observe that the dynamic range goes up to 127dB which is 63dB higher than that of CIS. However, because of the lower peak offered by an individual LDR, the overall peak of this HDR image is still lower than that of a CIS.

The result of this figure shows that we can trade-off the peak SNR and the dynamic range of a QIS by controlling the exposure pattern, e.g., using fewer but longer exposures, or more but shorter exposures. This flexibility can be of importance for various imaging applications.

\section{HDR Reconstruction for QIS}\label{sec:HDR_rec}
The significance of Theorem~\ref{thm:mub_sigb} is two-fold. In the previous section, we have seen how it informs us of the SNR and hence the dynamic range of QIS. In this section, we use Theorem~\ref{thm:mub_sigb} to derive an optimal linear HDR reconstruction algorithm.

\subsection{Exposure bracketing}
Before we discuss the problem formulation, we should first comment on how a conventional CIS performs HDR reconstruction. To a large extent, conventional HDR methods are based on the concept of exposure bracketing \cite{mann1994beingundigital}. Given a stack of differently exposed images, we construct a \emph{linear combination} of the images to create the final image. Putting this mathematically, if we denote $\mY[1],\ldots,\mY[N]$ as a sequence of $N$ differently exposed images, then the HDR image $\vlambdahat$ is
\begin{equation}
\vlambdahat = \sum_{n=1}^N \vw[n] \odot \mY[n],
\end{equation}
where $\odot$ denotes the element-wise multiplication, and $\{\vw[n]\}$ is a sequence of weight vectors satisfying the constraint that $\sum_{n=1}^N \vw[n] = \vone$. Because the reconstructed image $\vlambdahat$ is the linear combination of the input frames, we call such an exposure bracketing technique a \emph{linear} reconstruction method.  We follow the literature by deriving the theoretical results for static scenes. We leave the methods for dynamic scenes to future work.

\subsection{Optimal weights for QIS}
Without loss of generality let us assume that the QIS has acquired a stack of frames as given by \eref{eq: calY}, where the frames are grouped into $M$ different index sets of exposures $E_1,\ldots,E_M$. For example, the set $E_1$ contains all the indices of the frames that have used the exposure $\Delta_1$. For simplicity, we further assume that each $E_m$ contains $K$ frames. So for $M$ exposures, each having $K$ frames, the total number of frames is $N = KM$.

To make the notation simple we focus only on one pixel. Thus, the vectorized equation can be simplified to a scalar equation. Moreover, we assume that $\lambda(t) = \lambda$ for all $t$ because the scene is static. Under these assumptions, the average number of photons obtained by each exposure $\Delta_m$ is
\begin{equation}
    \theta[m] = \Delta_m \lambda,
\end{equation}
Intuitively, since the flux $\lambda$ is constant, the average number of photons is proportional to the exposure time $\Delta_m$.

Following the QIS model, each $\theta[m]$ will generate $K$ observations $Y[n_1],\ldots,Y[n_K]$. Depending on the ADC, each $Y[n]$ can be a one-bit or a multi-bit Poisson random variable. The mean and variance of each $Y[n]$ are respectively defined as
\begin{equation}
\mu_Y[m] \bydef \E[Y[n]], \;\;\;\text{and}\;\;\; \sigma_Y^2[m] \bydef \Var[Y[n]],
\end{equation}
for $n \in E_m$, where $m = 1,\ldots,M$. Essentially, this equation says that when we divide the exposures into $M$ groups, we have $M$ different means and variances.

Note that $\mu_Y[m]$ is a function of $\theta[m]$; $Y[n]$ is the truncated Poisson random variable according to the QIS model, and so $\mu_Y[m]$ must be a function of the underlying average photon count $\theta[m]$. Denoting $\mu_Y[m] = f(\theta[m])$ for some function $f$, it holds that $\theta[m] = f^{-1}(\mu_Y[m])$. For example, if the ADC is 1-bit, then
\begin{equation*}
\mu_Y[m] = 1-e^{-\theta[m]} \bydef f(\theta[m]),
\end{equation*}
and so $f^{-1}(\mu_Y[m]) = -\log(1-\mu_Y[m])$. As mentioned in \cite{Chan16}, this can be regarded as a tone-mapping.

As far as estimation is concerned, we reconstruct a low dynamic range (LDR) image from a stack of $K$ frames of the same exposure. We thus define the sum as
\begin{equation}\label{eqn:LDR_recon}
    S[m] \bydef \frac{K}{\Delta_m} f^{-1}\left( \frac{1}{K} \sum_{n \in E_m}Y[n]\right).
\end{equation}
Here, the quantity inside $f^{-1}$ is the average frames. $f^{-1}$ resolves the tone-mapping. The normalization $1/\Delta_m$ ensures that $S[m]$ is properly scaled with respect to the exposure time.

To construct the HDR image, we consider using a linear combination scheme by defining
\begin{equation}
    \widehat{\lambda} = \sum_{m=1}^M w[m] S[m],
    \label{eq: thetahat}
\end{equation}
where $w[m] \in \R$ is a weight satisfying the property that $\sum_{m=1}^M w[m] = 1$. Because of the weighted averaging instead of a simple sum, the exposure referred SNR for this estimator $\widehat{\lambda}$ takes a generalized form of \eref{eqn:SNRH_defn}. Specifically, since each exposure has $K$ frames, the signal in the numerator of \eref{eqn:SNRH_defn} is
\begin{equation*}
    \text{signal}^{\text{HDR}} = K\lambda.
\end{equation*}
The denominator of \eref{eqn:SNRH_defn}, which is the exposure-referred noise, becomes
\begin{equation}
    \text{noise}^{\text{HDR}} = \sqrt{\sum_{m=1}^M \left(\frac{w[m]}{\Delta_m}\right)^2 \sigma_{\text{H}}^2[m]},
\end{equation}
where $\sigma_{\text{H}}[m]$ is the exposure-referred noise standard deviation of the $m$-th exposure \footnote{$\sigma_\text{H}[m]$ can be obtained by taking the variance of $\lambdahat$. The derivative appears as a result of applying the delta method to $f^{-1}(S[m])$.}:
\begin{equation}
\sigma_\text{H}[m] = \sqrt{K} \sigma_Y[m] \cdot \frac{d\theta[m]}{d\mu_Y[m]}.
\label{eq: sigmaH, recon}
\end{equation}
Here, $\sigma_Y[m]$ follows from \eref{eq: Var[Y]}, and $\frac{d\theta[m]}{d\mu_Y[m]}$ follows from \eref{eq: dmu/dtheta}, where for each exposure $m$ there is a different $\sigma_Y[m]$ and $\frac{d\theta[m]}{d\mu_Y[m]}$. Taking the ratio between $\text{signal}^{\text{HDR}}$ and $\text{noise}^{\text{HDR}}$ gives us the overall SNR of the HDR image:
\begin{equation}
    \text{SNR}_{\text{H}}^{\text{HDR}}
    = \frac{K \lambda}{\sqrt{\sum_{m=1}^M \left(\frac{w[m]}{\Delta_m}\right)^2 \sigma_{\text{H}}^2[m]}}.
\end{equation}

The optimization problem is to find the optimal weights $w[1],\ldots,w[M]$ such that $\text{SNR}_{\text{H}}^{\text{HDR}}$ is maximized. This gives the following constrained problem:
\begin{eqnarray}
\label{eq: P1}
\begin{aligned}
\maximize{w[1],\ldots,w[M]} \quad & \frac{K \lambda}{\sqrt{\sum_{m=1}^M \left(\frac{w[m]}{\Delta_m}\right)^2 \sigma_{\text{H}}^2[m]}}\\
\text{subject to}           \quad & \sum\limits_{m=1}^M w[m] = 1, \;\text{and}\; w[m]\geq 0.
\end{aligned}
\end{eqnarray}
To specify the solution of this optimization problem, we define the $m$-th SNR as
\begin{equation}
    \text{SNR}_{\text{H}}[m] \bydef \frac{\theta[m]}{\sigma_{\text{H}}[m]} = \frac{\Delta_m \lambda}{\sigma_{\text{H}}[m]}.
        \label{eq: SNR H recon}
\end{equation}
With this definition, we can determine the solution.
\begin{theorem}
\label{thm: Thm2}
The optimal weights $w[1],\ldots,w[M]$ which solves the optimization problem \eref{eq: P1} is given by
\begin{equation}
    w[m] = \frac{\text{SNR}_\text{H}^2[m]}{\sum_{m=1}^M \text{SNR}_{\text{H}}^2[m]},
    \label{eq: thm 2 weight}
\end{equation}
where $\text{SNR}_{\text{H}}[m]$ is defined by \eref{eq: SNR H recon}.
\end{theorem}
\begin{proof}
See supplementary document \cite{gnanasambandam2020hdr}.
\end{proof}

\subsection{Comparison with CIS}
It is important to understand why a CIS-based reconstruction such as \cite{granados2010optimal} does not work for QIS. A CIS assumes a linear sensor response until the photon level reaches the full-well capacity, whereas QIS assumes a nonlinear response. The linear response of a CIS implies that \emph{before} saturation we have
$\mu_Y[m] = \theta[m]$ so that $d\theta[m]/d\mu_Y[m] = 1$ in \eref{eq: sigmaH, recon}, and \emph{after} saturation, we have that $\mu_Y[m] = L$ where $L$ is the full-well capacity and so $d\theta[m]/d\mu_Y[m] = \infty$. For $K$ frames, each with an exposure $\Delta_m$, the exposure-referred SNR is
\begin{align}
    \sigma_{\text{H}}[m]
    &= \sqrt{K} \cdot \sigma_Y[m] \cdot \frac{d\theta[m]}{d \mu_Y[m]} \notag \\
     &=
    \begin{cases}
    \sqrt{K} \sqrt{\Delta_m \lambda}, &\quad\mbox{if}\;\; \Delta_m \lambda < L,\\
    \infty, &\quad\mbox{if}\;\; \Delta_m \lambda \ge L.
    \end{cases}
\end{align}
Substituting this into $\text{SNR}^{\text{HDR}}$, we show that for CIS,
\begin{equation}
    \text{SNR}_{\text{H}}^{\text{HDR}} = \frac{K\lambda}{\sqrt{\sum_{m=1}^M \left(\frac{w[m]}{\Delta_m}\right)^2 K \Delta_m \lambda \cdot \mathbb{I}\left\{ \Delta_m \lambda < L\right\}}},
\end{equation}
where $\mathbb{I}\{\cdot\} = 1$ if the argument is true, and is $\infty$ if the argument is false. Consequently, one can solve a similar optimization as we did to obtain the following weight
\begin{equation}
    w[m] =
    \frac{\Delta_m \cdot \mathbb{I}\left\{ \Delta_m \lambda < L\right\} }{\sum_{m=1}^M \Delta_m \cdot \mathbb{I}\left\{ \Delta_m \lambda < L\right\}}.
\end{equation}
Therefore, as long as the pixels are not saturated for each exposure, the weight is linear with respect to the exposure time $\Delta_m$. This should be intuitive, because when the pixels are not saturated, longer exposure time gives higher SNR and so it should be weighted more. If a pixel becomes saturated, then the SNR will drop abruptly so that the corresponding exposure is invalidated.

The analysis here shows why a CIS-based reconstruction method does \emph{not} apply to QIS. QIS simply does not have the linear response as CIS does. As a result, the optimal linear reconstruction method for QIS given by Theorem~\ref{thm: Thm2} is not transferable to CIS, and vice versa.

\subsection{Reconstruction algorithm}
Theorem~\ref{thm: Thm2} suggests a method to construct an HDR image. The idea is that if we knew $\text{SNR}_{\text{H}}[m]$, then the weight is given according to \eref{eq: thm 2 weight}. Substituting the weight into \eref{eq: thetahat} will give us the estimate.

In practice, however, since we do not know $\lambda$, we need to estimate $\text{SNR}_{\text{H}}[m]$. The estimation is based on an iterative procedure. Denoting $w^{k}[m]$ as the weight at the $k$-th iteration, and $\lambdahat^{k}$ as the estimated HDR pixel in the $k$-th iteration, the iterative procedure is given by two steps:
\begin{align*}
\lambdahat^{k+1} &= \sum_{m=1}^M w^{k}[m] S[m],\\
w^{k+1}[m]       &= \frac{(\text{SNR}^{k+1}_{\text{H}}[m])^2}{\sum_{m=1}^M (\text{SNR}^{k+1}_{\text{H}}[m])^2},
\end{align*}
where $\text{SNR}^{k+1}_{\text{H}}[m]$ is evaluated based on \eref{eq: SNR H recon}, and the exposure referred noise $\sigma_{\text{H}}[m]$, which is a function of $\lambdahat$, is updated using Theorem~\ref{thm:mub_sigb}. The algorithm is summarized in Algorithm~\ref{algo:HDR_rec}.

\begin{algorithm}
 1. Acquire QIS frames $Y[n]$, $n = 1,\hdots N$.\\
 2. Obtain $M$ LDR images according to \eref{eqn:LDR_recon}.\\
 3. Initilialize $w^0[m] = 1/M, \forall m = 1, \hdots M$. \\
 4. Estimate the HDR image $\lambdahat^{k}$ according to \eref{eq: thetahat}.\\
 5. Update $\text{SNR}_\text{H}^{k}[m]$ according to \eref{eq: SNR H recon}.\\
 6. Update weights $w^k[m]$ according to \eref{eq: thm 2 weight}.\\
 7. Repeat 4,5,6 till convergence.\\
\caption{HDR Image Reconstruction \label{algo:HDR_rec}}
\end{algorithm}

\subsection{Practical considerations}
\textbf{Denoising.} The proposed HDR reconstruction method does not include any pre-processing of the input LDR images. In practice, it may be desired to perform some degree of denoising using simple methods such as the one introduced in \cite{Chan16}. The denoising is particularly useful when the number of frames is low. HDR denoising itself is an open problem. We leave the problem on denoising+HDR reconstruction as future work.

\textbf{Look up tables.} The proposed reconstruction method requires calculating the exposure-referred SNR for every pixel at the exposure period. This is computationally very expensive. However, we notice that the exposure-referred SNR is a function of the mean number of photons collected by the sensor at each frame. It is therefore possible to construct a look-up table to store the values by discretizing the mean signal levels. During the computation, one can refer to the look-up table when calculating $\text{SNR}_\text{H}[m]$.

\textbf{Dynamic Scenes.} The optimal reconstruction scheme presented in this paper is analogous to the optimal linear schemes in the conventional CIS-based HDR problems \cite{kirk2006noise,robertson2003estimation,granados2010optimal}. Thus, by design, the method is used for static scenes. We acknowledge the importance of HDR imaging for dynamic scenes. However, in the presence of shot noise and motion, the reconstruction problem becomes substantially harder. Several methods have demonstrated the feasibility of handling photon limited data and motion, e.g., \cite{Ma_SIGGRAPH20,yiheng_vladlen,gyongy2018single}. Adding exposure bracketing to these problems is an important future problem.

\begin{figure}[h]
    \centering
    \includegraphics[width = \linewidth]{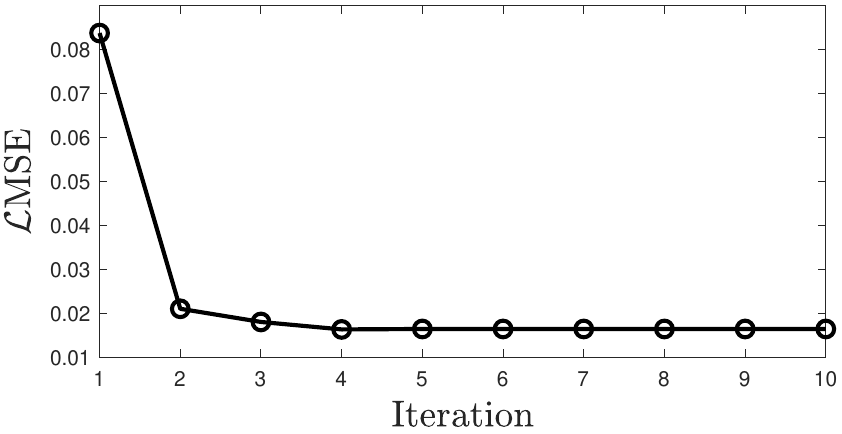}
    \vspace{-2.0ex}
    \caption{Number of iterations for the proposed algorithm to converge. We use three different integration times and 100 frames per integration time and use the proposed HDR reconstruction method.. The figure shows that $\calL$MSE converges after 5 iterations.}
    \label{fig:iterations}
\end{figure}

\textbf{Number of iterations.} The proposed reconstruction algorithm is iterative. In \fref{fig:iterations}, we plot the mean squared error in log scale ($\calL$MSE) as used in \cite{eilertsen2017hdr} between the reconstructed image and the ground truth image after each iteration. We use the ``aisle'' image from the Stanford HDR image dataset \cite{xiao2002high} for simulating the QIS data for this experiment. We use three different integration times and 100 frames per integration time and use the proposed HDR reconstruction method. We observe that $\calL$MSE converges after 5 iterations. We notice similar results with multiple images, different integration times and different number of frames.

\begin{figure*}[!t]
\centering
\begin{tabular}{c}
\includegraphics[width=0.95\linewidth]{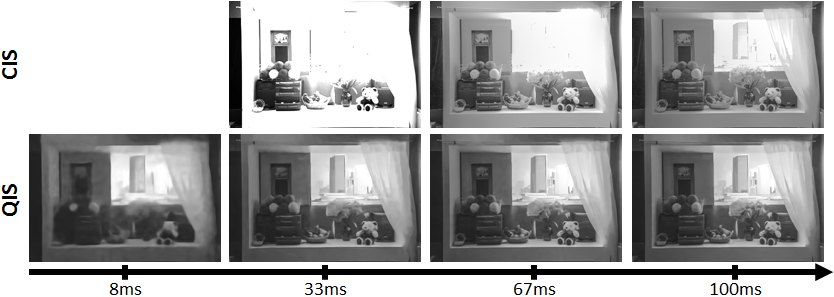}
\end{tabular}
\caption{\textbf{Comparing CIS and QIS for HDR imaging}. The CIS image is constructed from three frames, each with an exposure of $33$ ms, $3.3$ ms, and $0.33$ ms, respectively. The QIS image is constructed from a set of exposures $1.1$ ms, $0.11$ ms, and $0.011$ ms. The CIS is assumed to have a full well capacity of 4000 electrons. The number of 1-bit QIS frames is 30 times that of CIS so that the overall integration time for CIS and QIS are equal. The timestamps shown at the bottom of the figure are the overall integration time to capture all the exposures. Note that for short integration, e.g., 33ms or lower, QIS offers substantially better image reconstruction.}
\label{fig:new_plot_demo}
\end{figure*}

\section{Experiments}
In this section, we report the experimental results. Our results can be divided into two parts: (i) Comparing CIS with QIS for HDR imaging; (ii) Comparing the optimal HDR reconstruction algorithm and the existing methods.

\subsection{Comparing CIS with QIS for HDR imaging}
The first experiment evaluates the significance of QIS compared to CIS for HDR imaging. Some of the results have already been shown. We summarize them here:
\begin{itemize}
\setlength\itemsep{0em}
    \item \fref{fig:DR_cpmpare} illustrates the dynamic range that can be offered by one CIS frame (in 1ms), and that offered by multiple QIS frames of different bit-depths (within the same 1ms). Our result shows that CIS saturates whereas QIS does not. \item \fref{fig:tradeoff} shows the theoretical dynamic range of CIS and QIS. We observe that a single QIS exposure has a dynamic range of 10dB higher than that of a CIS. Fusing multiple exposures will widen the gap even further.
\end{itemize}
In addition to these results, we show in \fref{fig:new_plot_demo} a visual comparison between a CIS and a QIS. This experiment considers the practical frame rate limit of a QIS, which was assumed to be 1000 frames per second according to \cite{Ma:17}. This is approximately 30 times faster than a standard CIS operating at 30 frames per second \cite{mclernon2012canon}. While there exists even faster QIS prototypes (e.g., \cite{bruschini2018monolithic}), \fref{fig:new_plot_demo} shows that with 1000 fps, QIS already offers an advantage over the CIS.

To conduct this experiment, we simulate 30 QIS frames for every CIS frame. The bit-depth of the QIS is 1-bit. Among the QIS exposures, we consider the multi-exposure scheme consisting of integration times $1.1$ ms, $0.11$ ms, and $0.011$ ms. For CIS, we use integration times $33$ ms, $3.3$ ms, and $0.33$ ms. The CIS is assumed to have a full well capacity of 4000 electrons.  We use the proposed HDR reconstruction method for obtaining the QIS HDR image and \cite{granados2010optimal} for the CIS HDR image. Notice that  CIS produces good quality images with limited dynamic range initially, and the dynamic range improves over time. Compared to this, the QIS can produce images with a larger dynamic range at only a fraction of the time taken by the CIS to produce its first frame. Although the images are noisy initially, the quality gets better over time. At 100 ms the quality and the dynamic range of both the QIS and CIS images are about the same. But, when the total time taken reduces, QIS offers a higher dynamic range than the CIS.

\begin{figure}[t]
    \centering
\begin{tabular}{cc}
\hspace{-2.0ex}
\includegraphics[width=0.47\linewidth]{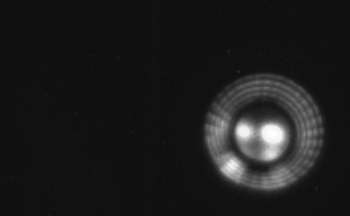} &
\hspace{-2.0ex}
\includegraphics[width=0.47\linewidth]{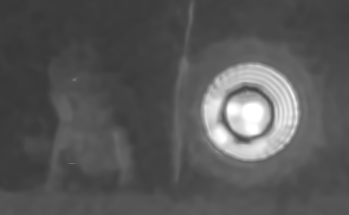} \\
CIS $N=1$ & 1-bit QIS $N=20$
\end{tabular}
\caption{\textbf{Comparing CIS and QIS}. In this real experiment, we use a commercially available CIS and compare it with a prototype QIS. Within a fixed integration time, CIS only captures one frame whereas QIS has captured multiple frames of different exposures.}
\label{fig:real_QIS_CIS}
\end{figure}

In \fref{fig:real_QIS_CIS}, we compare QIS and CIS using real data. We collect a total of $N=20$ 1-bit QIS frames, with $K = 10$ frames at 2 different integration times of $50 \mu s$, $1000 \mu s$. We compare this to a CIS image obtained using e-con System's e-CAM40\_CUMI4682\_MOD camera module which uses OmniVision's OV4682 image sensor. \fref{fig:real_QIS_CIS} shows a clear distinction between two sensors.

\begin{figure*}[ht]
\centering
\begin{tabular}{ccccc}
(a)& (b) \cite{dutton2018high} & (c) \cite{granados2010optimal} & (d) Ours & (e) Ground Truth \\
\includegraphics[width=0.0565\linewidth]{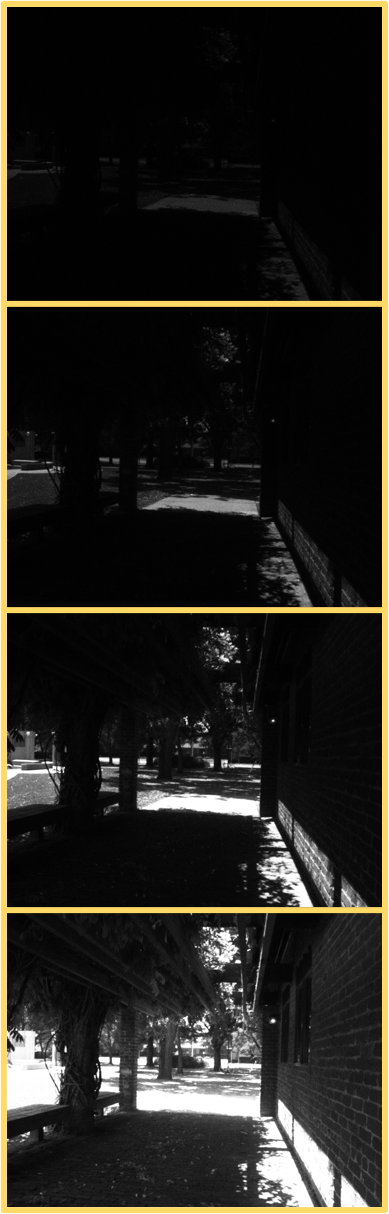} &
\hspace{-2.0ex}\includegraphics[width=0.22\linewidth]{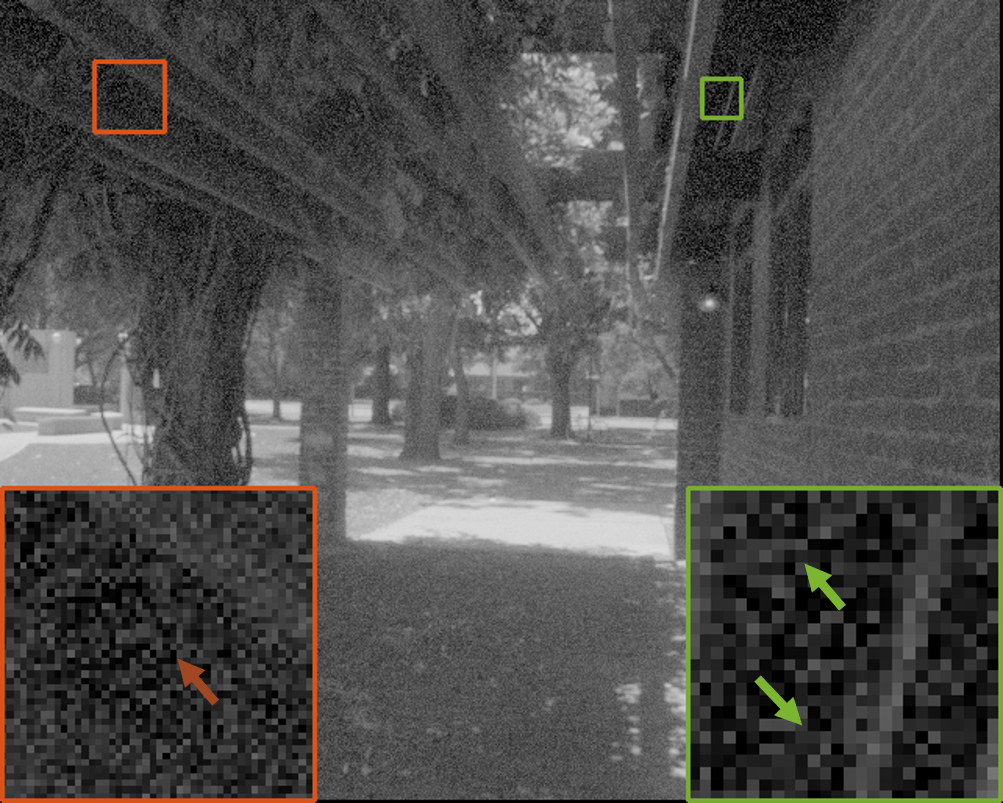} &
\hspace{-2.0ex}\includegraphics[width=0.22\linewidth]{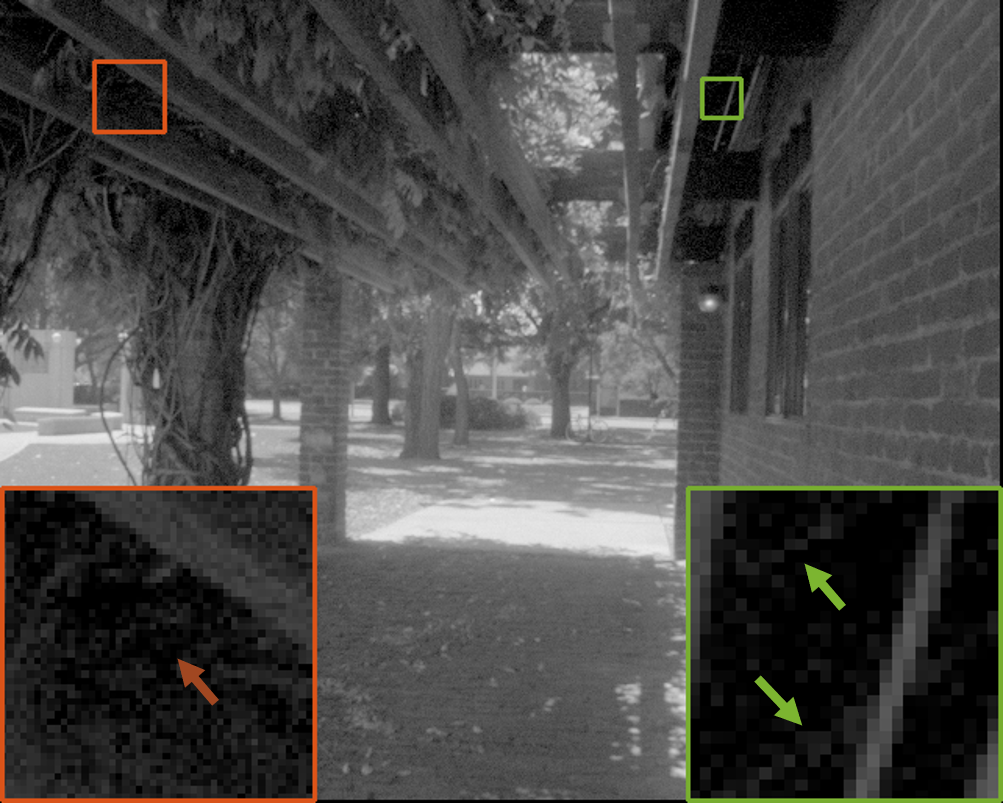} &
\hspace{-2.0ex}\includegraphics[width=0.22\linewidth]{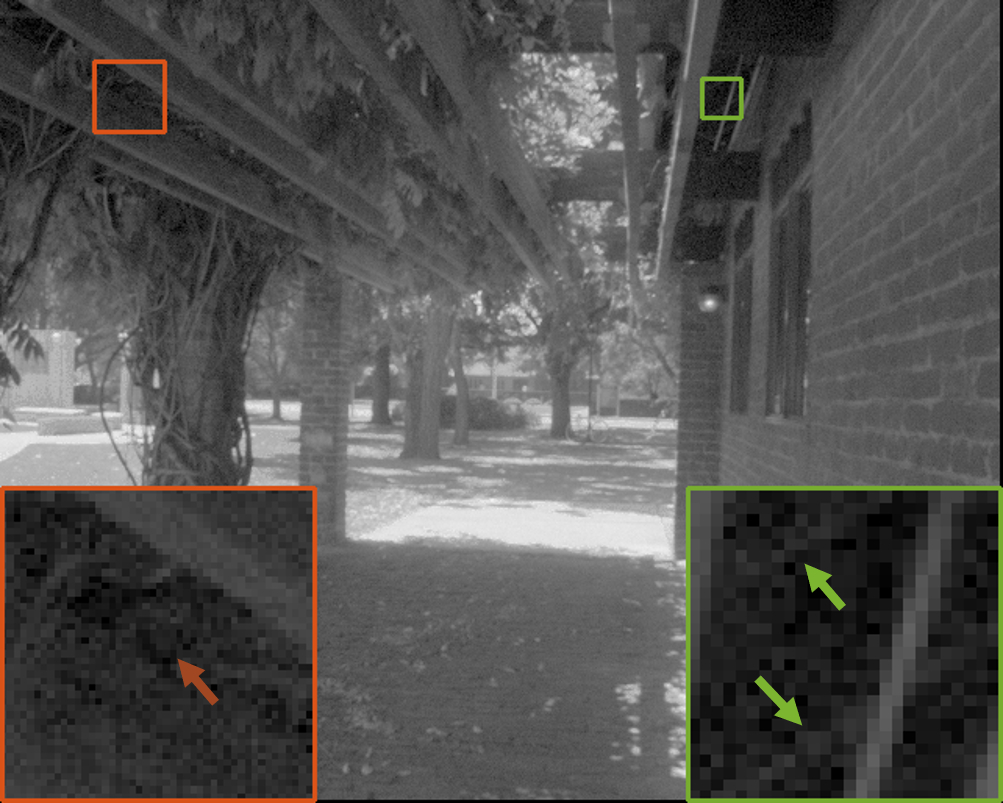} &
\hspace{-2.0ex}\includegraphics[width=0.22\linewidth]{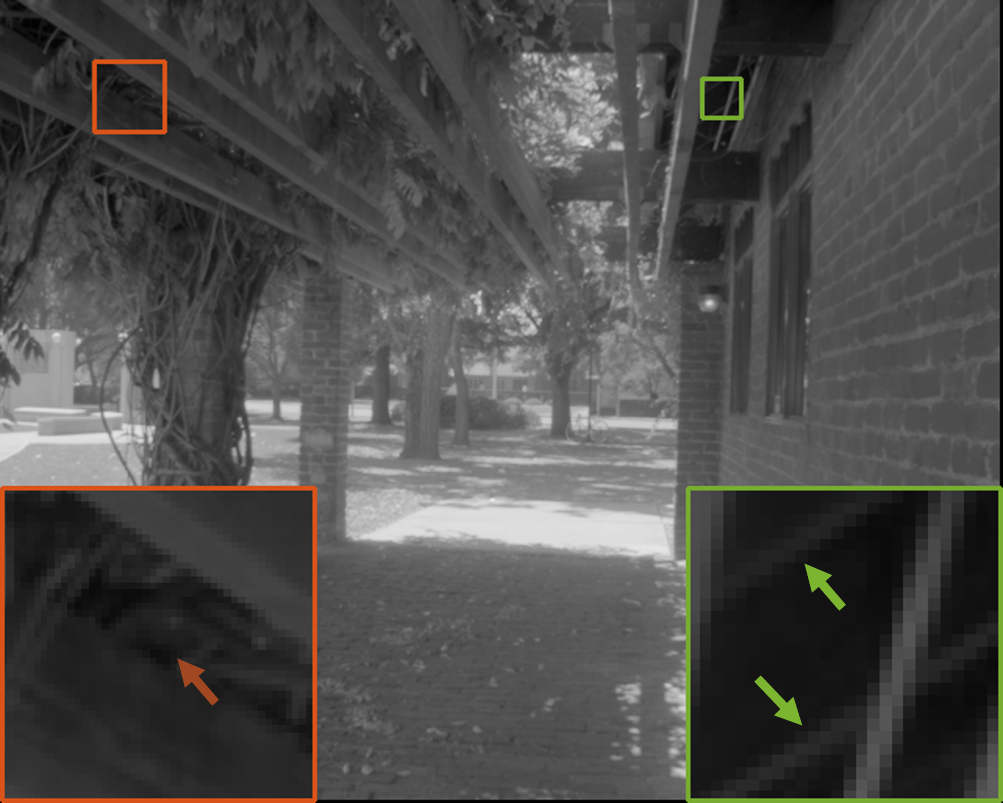} \\
&$\calL$MSE = 0.1310 & $\calL$MSE = 0.0.0776 & $\calL$MSE = 0.0653&
\vspace{-1ex}
\end{tabular}
\caption{\label{fig:algo_compare}\textbf{The HDR Reconstruction Algorithm - Synthetic Experiment}. A total of 400 3 bit frames were simulated at 4 different integration times, with 100 frames at each integration time. We can clearly see that the HDR images obtained in (d) using the proposed method are closer to the ground truth than the other two methods. The images are displayed on the log-scale. $\calL$MSE is the mean squared error measure in $\log$-scale. (Images courtesy : \cite{xiao2002high})}
\end{figure*}

\begin{figure*}[t]
\centering
\begin{tabular}{ccc}
\includegraphics[width=0.32\linewidth]{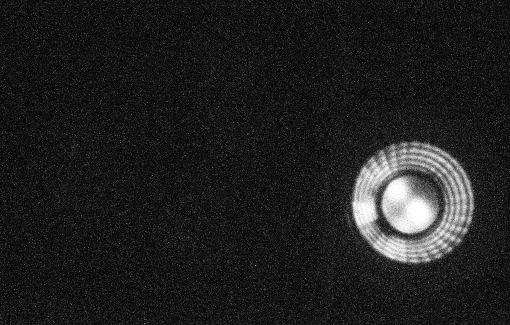}&
\hspace{-2.0ex}\includegraphics[width=0.32\linewidth]{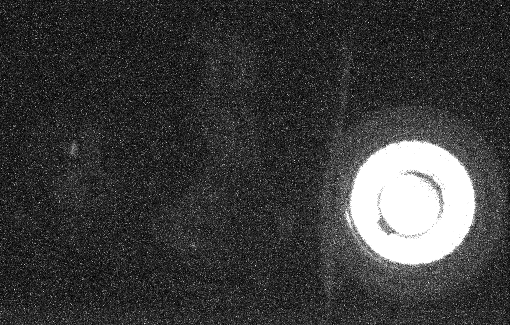}&
\hspace{-2.0ex}\includegraphics[width=0.32\linewidth]{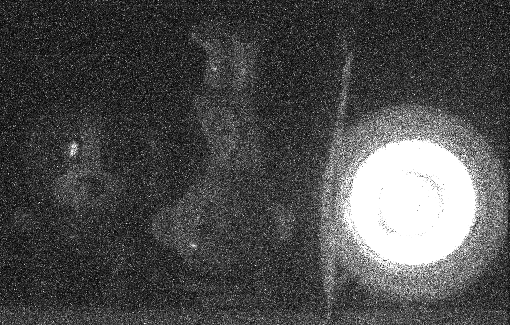}\\
$\Delta = 75 \mu s$, 1 frame & $\Delta = 375 \mu s$, 1 frame & $\Delta = 1875 \mu s$, 1 frame\\
\includegraphics[width=0.32\linewidth]{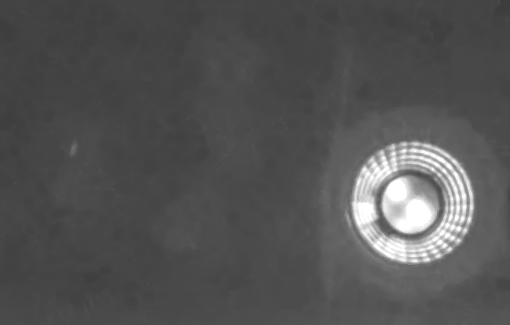}&
\hspace{-2.0ex}\includegraphics[width=0.32\linewidth]{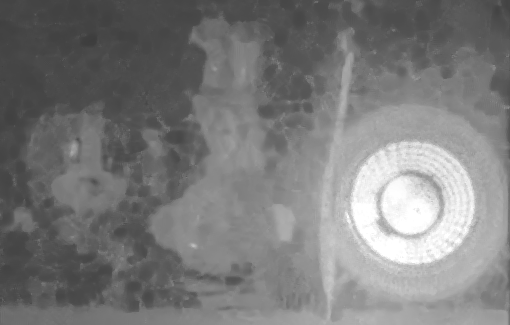}&
\hspace{-2.0ex}\includegraphics[width=0.32\linewidth]{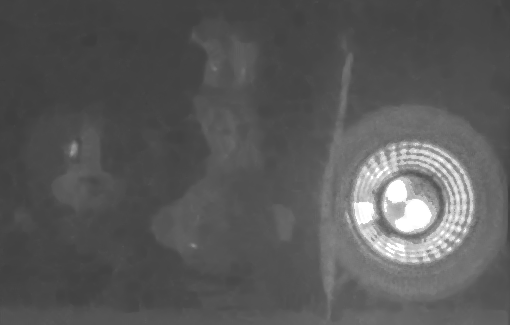}\\
Reconstruction using 30 frames \cite{dutton2018high} & Reconstruction using 30 frames  \cite{granados2010optimal} & Reconstruction using 30 frames, Ours
\end{tabular}
\vspace{-2ex}
\caption{\textbf{The HDR Reconstruction Algorithm - Real Experiment} In this experiment, we collect 10 QIS frames, each at 3 different exposures - $75 \mu s$, $375 \mu s$, and $1875 \mu s$ in 3-bit modes. The result shows the advantage of the proposed HDR reconstruction methods over the other two methods.}
\label{fig:realdata}
\end{figure*}

\subsection{Reconstruction algorithm}
The second experiment evaluates the optimal reconstruction scheme. While we acknowledge the promising results of deep neural network solutions, in this paper we compare with two deterministic schemes \cite{dutton2018high} and \cite{granados2010optimal} for three reasons:
\begin{itemize}
    \item The objective of this paper is not to compete with state-of-the-art HDR image reconstruction algorithms that are customized for CIS. Moreover, there does not exist QIS datasets for us to conduct a fair comparison.
    \item Among the deterministic methods, \cite{granados2010optimal} is theoretically optimal for CIS. No other linear methods can achieve better results. We compare with this method to show that CIS methods cannot be translated to QIS.
    \item Among the QIS methods, \cite{dutton2018high} is one of the latest works in the literature. We compare with this method to show the effectiveness of our method.
\end{itemize}

We first evaluate the methods using the Stanford-HDR dataset \cite{xiao2002high} containing 88 HDR images. We normalize the images such that the $0.01 \leq \lambda \Delta \leq 8000, \text{if\;} \lambda \neq 0$ at every pixel. We simulate a total of $N = 3000$ 1-bit and 3-bit frames with with $K=1000$ frames each at 3 different integration times of $\Delta$, $\Delta/10$, and $\Delta/100$. We use the $\calL$MSE, PU-PSNR and PU-SSIM \cite{aydin2008extending}  as the metrics for comparison. $\calL$MSE measures the mean squared error (MSE) in log-scale. PU-PSNR and PU-SSIM calculate peak signal-to-noise ratio (PSNR) and structural similarity (SSIM) using a perceptually uniform (PU) encoding.   We compare the performance of the proposed HDR reconstruction method with reconstruction methods from \cite{dutton2018high} and \cite{granados2010optimal} in \tref{tab:stanford}, using the average $\calL\text{MSE}$, PU-PSNR, and PU-SSIM compared to the ground-truth for the three methods across the 88 HDR images. We see that the proposed method outperforms the two competing methods in all the three metrics that we have considered, in both single bit and three bit modes.

\begin{table*}[t]
\caption{\label{tab:stanford} Comparing the three HDR reconstruction methods. }
\centering

\begin{tabular}{cccc|ccc}
& \multicolumn{3}{c}{1 bit} & \multicolumn{3}{c}{3 bits} \\
 & Dutton et al. &  Granados et al. & & Dutton et al. &  Granados et al. &\\
Metric & \cite{dutton2018high} &  \cite{granados2010optimal} & Proposed & \cite{dutton2018high} &  \cite{granados2010optimal} & Proposed \\
\hline
$\calL$MSE & $11.25 \times 10^{-2}$ & $1.23 \times 10^{-2}$ & $\mathbf{ 0.61 \times 10^{-2}}$ & $10.02 \times 10^{-2}$ & $0.59 \times 10^{-2}$ & $\mathbf{0.49 \times 10^{-2}}$\\
PU-PSNR & 32.53 & 34.89 & \textbf{35.92} &  33.26 & 36.42 & \textbf{36.81} \\
PU-SSIM &0.9138 & 0.9822 & \textbf{0.9850} & 0.9345 & 0.9901 & \textbf{0.9912}\\
\hline
\end{tabular}
\end{table*}

In \fref{fig:algo_compare}, we visually compare the three methods. We use 3-bit images. 100 frames are collected at 4 different integration times, thus giving a total of 400 frames. These frames are then used to reconstruct the high dynamic range image. Notice that the proposed method outperforms \cite{dutton2018high} and \cite{granados2010optimal}, both visually and the in the $\calL \text{MSE}$ metric.

Next, we show the comparisons using real QIS data in \fref{fig:realdata}. We collect a total of $N=30$ frames of 3-bit QIS data, with $K = 10$ frames at 3 different integration times of $75 \mu s$, $375 \mu s$ and $1875 \mu s$. The scene consists of a bright light bulb on the right and two dark objects on the left. The three LDR images show different levels of saturation. We apply \cite{dutton2018high} and \cite{granados2010optimal} to the image stack and reconstruct a HDR image. We observe that the method by Dutton et al. \cite{dutton2018high} has a weak reconstruction of the darker regions since it provides equal weights to all the three integration times. The method by Granados et al. \cite{granados2010optimal} has better dynamic range but it also generates artifacts in the brighter regions. The proposed method, which is optimal for QIS, produces an HDR with fewer artifacts.

\begin{figure*}[t]
\centering
\begin{tabular}{cccccc}
\rotatebox{90}{\ \ \ \ \ \ \ \ \ \ \ \textbf{1 bit}} &
\includegraphics[width=0.07\linewidth]{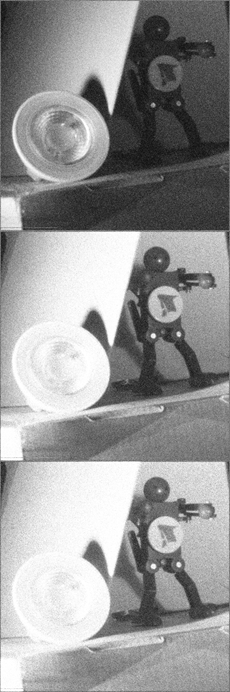}&
\hspace{-2.0ex}\includegraphics[width=0.21\linewidth]{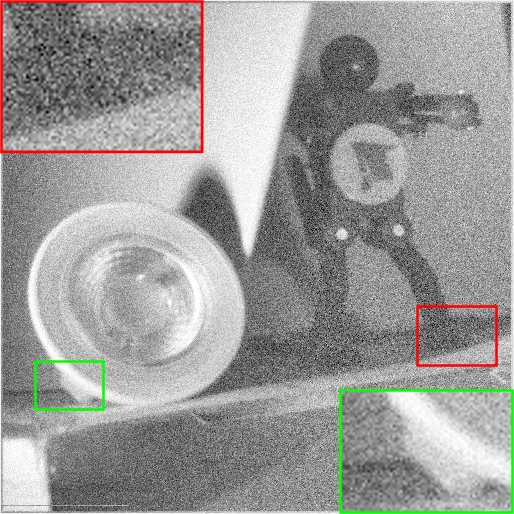}&
\hspace{-2.0ex}\includegraphics[width=0.21\linewidth]{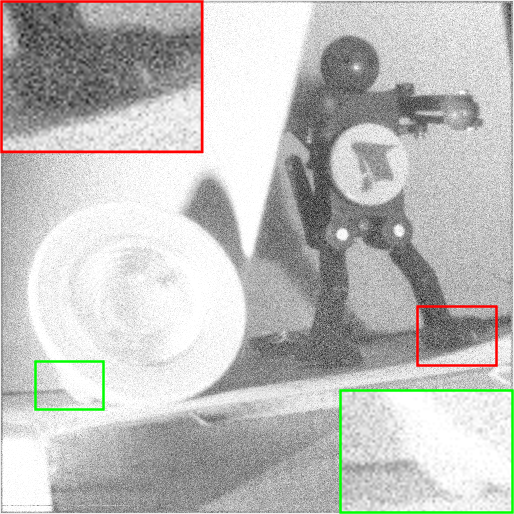}&
\hspace{-2.0ex}\includegraphics[width=0.21\linewidth]{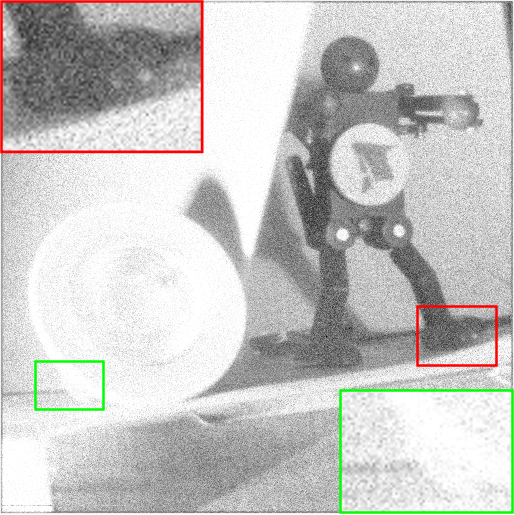} &
 \hspace{-2.0ex}\includegraphics[width=0.21\linewidth]{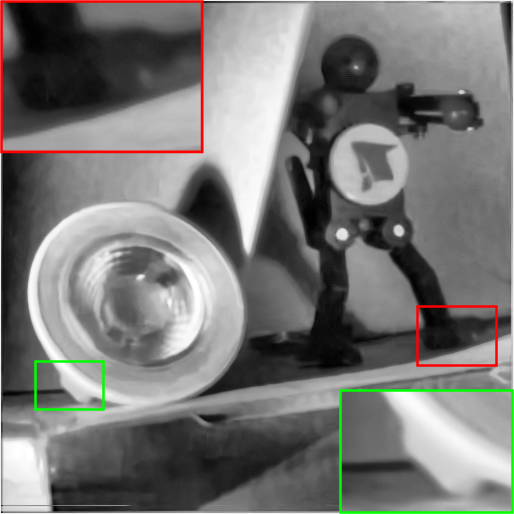}\\
&&$75 \mu s$ & $500 \mu s$ & $1100 \mu s$ & HDR image using\\
&&   $15\times 2 \times 2$       & $15\times 2 \times 2$            &$15\times 2 \times 2$                &proposed algorithm
\end{tabular}
\vspace{-1ex}
\caption{\textbf{Real Experiment}. In this experiment, we obtain a $K=15$ frames at 3 different integrationt times of $\Delta_1 = 75 \mu s$, $\Delta_2 = 575 \mu s$, and $\Delta_3 = 1175 \mu s$ . Spatial oversampling of $2\times2$ is used. The proposed HDR reconstruction algorithm is used to obtain the final HDR image. For display purpose, we use MATLAB's tonemap is re-scale the image intensity. The raw (un-normalized) images are show in the small insets on the first column.}
\label{fig:realdata_matlab}
\end{figure*}

Finally, we show the reconstruction results for an image containing more complex content. In \fref{fig:realdata_matlab}, we collect a total of $N=45$ frames, with $K=15$ frames each at 3 different integration times of $75 \mu s$, $575 \mu s$ and $1175 \mu s$. We use 1-bit QIS with a spatial oversampling factor of $2 \times 2$. The denoiser from \cite{Chan16} used for denoising the LDR image at each integration time, before using the proposed method for HDR reconstruction. As we can observe in the images, the short exposure captures the bright regions well but the image contains noise whereas the long exposure has better SNR but saturated at bright regions. The reconstructed HDR image has recovered the details and maintained the SNR.

\section{Conclusion}
Quanta image sensors can oversample a scene both spatially and temporally because of its ability to operate the sensor at a lower bit depth than a conventional CMOS Image Sensors. This ability, combined with lower read noise and dark current, provides QIS a unique advantage for capturing high dynamic range scenes. In this paper, we theoretically derive a closed-form expression for the signal-to-noise ratio of the QIS images. Using this result, we demonstrated the advantage of using a QIS over a CIS in terms of dynamic range and for high dynamic range imaging. We proposed a theoretically optimal high dynamic range reconstruction method. Using synthetic and real images, we demonstrated the advantage of QIS over CIS as well as the effectiveness of the proposed HDR reconstruction method.

\section*{Acknowledgement}
A shorter version of the paper was presented at the International Image Sensor Workshop 2019 \cite{gnanasambandamhigh}. The authors would like to express gratitude to Dr. Eric Fossum, and Dr. Jiaju Ma for providing suggestions to the IISW manuscript, and GigaJot Tech Inc. for providing the PathFinder QIS prototype camera for experiments. This work is supported, in part, by the National Science Foundation under grant CCF-1718007.

\section{Appendix}

\section*{ Proof of Theorem 1}
We first make the observation that
$$Y[n] = \text{ADC}\big(Z[n]\big) = \text{ADC}\big(\round{Z[n]}\big).$$

We introduce a new random variable variable $R[n] = \round{Z[n]}$. Now,
\begin{align}
\nonumber R[n] &= \round{Z[n]} \\
\nonumber &= \round{K[n] + \eta[n]} \\
\nonumber &= K[n] + \round{\eta[n]}.
\end{align}

The final step is possible because $K[n]$ is an integer. We introduce another new random variable $\gamma[n] = \round{\eta[n]}$. Now the pmf of $\gamma[n]$ is
\begin{equation}
\label{eqn:p_gamma}p_k = \mathbb{P}(\gamma[n] = k) = \int_{k-0.5}^{k+0.5} \frac{1}{\sqrt{2\pi \sigma_\text{read}^2} }e^{-\frac{x^2}{2\sigma_\text{read}^2}} dx.
\end{equation}

So, $R[n] = K[n] + \gamma [n]$, where
$$\mathbb{P}(K[n] = j) = \frac{e^{-\theta} \theta^j}{j!} \; \text{if} \;j \geq 0  $$ and $\gamma[n]$ is simulated according to \eref{eqn:p_gamma}. So,
$$\mathbb{P}(R[n] = j) = \sum\limits_{k=-\infty}^\infty p_k\cdot\mathbb{P}(K[n] = j-k).$$
Now, the probability mass function of $Y[n]$ is
$$
\mathbb{P}(Y[n] = i) =
\begin{cases}
\sum\limits_{j=-\infty}^0  \mathbb{P}(R[n] = j)     &\mbox{if}\; i = 0,\\
\mathbb{P}(R[n] = i)    &\mbox{if}\; 1 \leq i \leq L-1,\\
\sum\limits_{j=L}^\infty  \mathbb{P}(R[n] = j)     &\mbox{if}\; i = L,\\
0 &\; \text{otherwise}.
\end{cases}
$$

Now,
\begingroup
\allowdisplaybreaks
\begin{align}
\nonumber&\mathbb{E}(Y[n]) = \sum_{q=0}^L i\cdot\mathbb{P}(Y[n] = q) \\
\nonumber&= \sum_{q=1}^{L-1} q\cdot\mathbb{P}(Y[n] = q) + L.\mathbb{P}(Y[n] = L)\\
\nonumber&= \sum_{q=1}^{L-1} q\cdot\left(  \sum\limits_{k=-\infty}^\infty p_k\cdot\mathbb{P}(K[n] = q-k)\right) \\
\nonumber& + L\cdot\left( \sum\limits_{q=L}^\infty \sum\limits_{k = -\infty}^\infty p_k\cdot \mathbb{P}(K[n] = q-k)\right)\\
\nonumber&=  \sum_{k=-\infty}^{\infty} p_k\cdot\big(  \sum\limits_{q=1}^{L-1} q\cdot\mathbb{P}(K[n] =
q-k) \\
\nonumber& +\sum\limits_{q=L}^\infty L\cdot \mathbb{P}(K[n] = q-k)\big)\\
\nonumber& =\sum\limits_{q=1}^{L-1} q\cdot\mathbb{P}(K[n] = q) + L\cdot\sum\limits_{q=L}^\infty \mathbb{P}(K[n] = q) \\
\nonumber& \sum_{k=-\infty}^{\infty} p_k\cdot\big(  \sum\limits_{q=1}^{L-1} q \{ \mathbb{P}(K[n] =
q-k) - \mathbb{P}(K[n] =
q) \}\\
& +L\cdot \sum\limits_{q=L}^\infty \{\mathbb{P}(K[n] = q-k) - \mathbb{P}(K[n] =
q)\}\big) \label{eqn:thm1_proof1}
\end{align}
\endgroup
In \eref{eqn:thm1_proof1}, $\mathbb{E}(Y[n]) = \sum\limits_{q=1}^{L-1} q\cdot\mathbb{P}(K[n] = q) + L\cdot\sum\limits_{q=L}^\infty \mathbb{P}(K[n] = q) = \mathbb{E}(K[n])$, when the read noise $\sigma_\text{read} = 0$. The expression corresponding to this was derived in \cite{gnanasambandamhigh} as $\theta(\Psi_{L-1}(\theta )) + L(1 - \Psi_{L}(\theta))$. By re-arranging the rest of the terms and utilizing the fact that $\Psi_q(\theta) = \sum\limits_{k=0}^{q-1} \frac{\theta^k e^{-\theta}}{k!}$ and $\mathbb{P}(K[n] = j) = \frac{e^{-\theta} \theta^j}{j!}$, we can obtain the expression for $\mu_Y = \mathbb{E}(Y[n])$. We can clearly see that all the terms in \eref{eqn:thm1_proof1} is differentiable. Thus, taking the derivative of \eref{eqn:thm1_proof1} w.r.t. $\theta$ gives us the expression for $\frac{d \mu_Y}{d \theta}$

The expression for $\sigma_Y^2$ can also be calculated by following similar steps as above.

\section*{ Proof of Theorem 2}
The optimization problem is
\begin{eqnarray}
\label{opt:orig_proof}
\begin{aligned}
\maximize{w[1],\ldots,w[M]} \quad & \frac{K \lambda}{\sqrt{\sum_{m=1}^M \left(\frac{w[m]}{\Delta_m}\right)^2 \sigma_{\text{H}}^2[m]}}\\
\text{subject to}           \quad & \sum\limits_{m=1}^M w[m] = 1, \;\text{and}\; w[m]\geq 0.
\end{aligned}
\end{eqnarray}

Using a lagrange multiplier $\alpha$, we can re-write the optimization problem as
\begin{eqnarray}
\label{opt:changed_1}
\begin{aligned}
\min_{w_{i,j}} \quad & \sum\limits_{m=1}^M \left(w[m]\right)^2 \left(\frac{\sigma_\text{H}^2[m]}{{{\Delta_m}}}\right)^2 + \alpha\left( \sum\limits_{m=1}^M w[m] - 1\right)\\
\text{subject to}           \quad & \sum\limits_{m=1}^M w[m] = 1, \;\text{and}\; w[m]\geq 0.
\end{aligned}
\end{eqnarray}
 Solving this optimization problem, we get $$w[m] = \frac{\left(\frac{\Delta_m}{\sigma_H[m]}\right)^2}{\sum\limits_{k=1}^M\left(\frac{\Delta_k}{\sigma_H[k]}\right)^2}$$

 Comparing this result with the expression for $\text{SNR}_\text{H}$, we can obtain the necessary expression.

\bibliography{ref}
\bibliographystyle{ieeetr}

\end{document}